\documentclass[%
superscriptaddress,
 preprint,
 showpacs,preprintnumbers,
 nofootinbib,
 amsmath,amssymb,
aps,
]{revtex4-1}

\usepackage{graphicx}
\usepackage{dcolumn}
\usepackage{bm}
\usepackage{graphicx}
\usepackage{epstopdf}

\begin{document}


\title{Discrete Boltzmann modeling of Rayleigh-Taylor instability in two-component compressible flows}

\author{Chuandong Lin}
\email{chuandonglin@163.com}
\affiliation{Center for Combustion Energy; Key Laboratory for Thermal Science and Power Engineering of Ministry of Education, Department of Thermal Engineering, Tsinghua University, Beijing 100084, China}
\affiliation{State Key Laboratory for GeoMechanics and Deep Underground Engineering, China University of Mining and Technology, Beijing 100083, China}
\affiliation{College of Mathematics and Informatics $\&$ FJKLMAA, Fujian Normal University, Fuzhou 350007, China}

\author{Aiguo Xu}
\email{Xu\_Aiguo@iapcm.ac.cn}
\affiliation{Laboratory of Computational Physics, Institute of Applied Physics and Computational Mathematics, P. O. Box 8009-26, Beijing 100088, China}
\affiliation{Center for Applied Physics and Technology, MOE Key Center for High Energy Density Physics Simulations, College of Engineering, Peking University, Beijing 100871, China}

\author{Guangcai Zhang}
\email{zhang\_guangcai@iapcm.ac.cn}
\affiliation{Laboratory of Computational Physics, Institute of Applied Physics and Computational Mathematics, P. O. Box 8009-26, Beijing 100088, China}

\author{Kai Hong Luo}
\email{K.Luo@ucl.ac.uk}
\affiliation{Center for Combustion Energy; Key Laboratory for Thermal Science and Power Engineering of Ministry of Education, Department of Thermal Engineering, Tsinghua University, Beijing 100084, China}
\affiliation{Department of Mechanical Engineering, University College London, Torrington Place, London WC1E 7JE, UK}

\author{Yingjun Li}
\email{lyj@aphy.iphy.ac.cn}
\affiliation{State Key Laboratory for GeoMechanics and Deep Underground Engineering, China University of Mining and Technology, Beijing 100083, China}

\date{\today}

\begin{abstract}

A discrete Boltzmann model (DBM) is proposed to probe the Rayleigh-Taylor instability (RTI) in two-component compressible flows. Each species has a flexible specific heat ratio and is described by one discrete Boltzmann equation (DBE). Independent discrete velocities are adopted for the two DBEs. 
The collision and force terms in the DBE account for the molecular collision and external force, respectively. Two types of force terms are exploited. In addition to recovering the modified Navier-Stokes equations in the hydrodynamic limit, the DBM has the capability of capturing detailed nonequilibrium effects. Furthermore, we use the DBM to investigate the dynamic process of the RTI. The invariants of tensors for nonequilibrium effects are presented and studied. For low Reynolds numbers, both global nonequilibrium manifestations and the growth rate of the entropy of mixing show three stages (i.e., the reducing, increasing, and then decreasing trends) in the evolution of the RTI. On the other hand, the early reducing tendency is suppressed and even eliminated for high Reynolds numbers. Relevant physical mechanisms are analyzed and discussed. 

\end{abstract}


\maketitle


\section{Introduction}

Rayleigh-Taylor instability (RTI) occurs when a heavy fluid is accelerated or supported by a light one in a force field \cite{Rayleigh1882,Taylor1950}. It is a fundamental and ubiquitous fluid phenomenon in nature, as well as science and engineering \cite{Chandrasekhar1968}. In fact, it covers a wide range of applications, from microscopic to macroscopic levels, such as inertial confinement fusion, astrophysics, atmospheric science, oceanography, combustion, etc. Extensive efforts have been devoted to theoretical \cite{Goncharov2002,Wang2010,Tao2012}, experimental \cite{Lewis1950}, and computational studies of the phenomena \cite{Wei2012,Wang2016POP,XuLai2016,XuChen2016}. At present, it is still an open subject with many challenging issues, especially those relevant to hydrodynamic nonequilibrium (HNE) and thermodynamic nonequilibrium (TNE) phenomena \cite{Sagert2014,XuLai2016,XuChen2016}. To investigate those complex nonequilibrium manifestations, a rigorous approach is to employ the Boltzmann equation \cite{Qin2005,Qin2006,Zhang2008,Zhang2009,Zhang2011,Zhang2013,Zhang2016,Montessori2016} which describes the evolution of nonequilibrium statistical physical systems. However, solving the Boltzmann equation directly is computationally prohibitive. 

Based on the Boltzmann equation, the lattice Boltzmann method (LBM) has emerged as a feasible and versatile computational tool for describing the dynamics of complex systems, such as multiphase and/or multicomponent flows \cite{Succi-Book}. Over the last three decades, the LBM has been developed and modified with great enhancements in terms of precision and/or efficiency \cite{Benzi2011,Qin2015,Yeomans1998,Yeomans2014SM,Yeomans2014NP,Falcucci2007CCP,Falcucci2010SM,Falcucci2011CCP,Falcucci2010SAE,Falcucci2013JFM}. In 2010, Falcucci et al. \cite{Falcucci2010SAE} studied the cooperation between short and mid range attraction in LBMs with multi-range pseudo-potentials, which are applicable to phase separation at density ratios of liquid to vapor beyond $500:1$. Some interesting findings of complex phenomena (such as fluid instability, phase separating, stress-induced cavitation, spray formation and break-up) were obtained with the LBMs \cite{Falcucci2007CCP,Falcucci2010SM,Falcucci2011CCP,Falcucci2010SAE,Falcucci2013JFM}. In 2016, Li et al. \cite{LuoLi2016} provided a comprehensive review of the development of LBMs for thermofluids and energy applications. 

The RTI has also attracted a great deal of attention in the LBM field \cite{He1999POF,Zhang2000,Clark2003,Nourgaliev2003,Chiappini2010,Scagliarini2010,Biferale2011,LuoLi2012,Liu2013,Liang2014,Liang2016,Nie1998,Zu2013}. LBM methodologies for the RTI can be classified into two groups. One is the mono-component LBM for the RTI with a cold (hot) region on the top (bottom) half of a fluid system \cite{He1999POF,Zhang2000,Clark2003,Nourgaliev2003,Chiappini2010,Scagliarini2010,Biferale2011,LuoLi2012,Liu2013,Liang2014,Liang2016}. The other is the bi- or multi-component LBM for complex flows where temperatures are either identical or different for various species \cite{Nie1998,Zu2013}. Although the latter group has much fewer papers than the former, some successes have been achieved. For example, Nie et al. \cite{Nie1998} used the LBM for multi-component systems to simulate the RTI. The linear and mixing stages are simulated successfully. In 2013, based on the phase-field theory, Zu and He \cite{Zu2013} presented an LBM for incompressible binary fluids with density and viscosity contrasts. This model has the capability of simulating two-component systems with moderate density ratios. 

However, previous LBMs mainly work as merely solvers of traditional hydrodynamic equations, such as incompressible Navier-Stokes (NS) equations. To obtain a deeper insight into nonequilibrium flows, we need to investigate both the HNE and TNE effects beyond the NS equations. In addition to a better characterization of nonequilibrium flow state, the study on the TNE is helpful for understanding the nonlinear constitutive relation which significantly influences the physical accuracy of the hydrodynamic model. To this end, we resort to a  variant of the traditional LBM which is named the discrete Boltzmann model (DBM)  \cite{XuLai2016,XuChen2016,XuLin2014PRE,XuLin2015PRE,XuGan2015SM,XuLin2016CNF,XuZhang2016}. The DBM has been used to investigate various complex flows. Some new observations made by the DBM, for example, the fine physical structures of shock waves \cite{XuLin2014PRE}, have been confirmed and supplemented by molecular dynamics simulations \cite{Liu2016FOP,Liu2017PRE,Liu2017SS}. In 2016, Lai et al. \cite{XuLai2016} adopted the DBM to probe the effects of compressibility on RTI by inspecting the interplay between HNE and TNE phenomena. Almost at the same time, Chen et al. \cite{XuChen2016} utilized a multiple-relaxation-time DBM to investigate the viscosity, heat conductivity, and Prandtl number effects from macroscopic and nonequilibrium viewpoints. The two DBMs \cite{XuLai2016,XuChen2016} are only applicable to single component fluids where heavy (light) medium has low (high) temperature. For the sake of investigating the RTI in more common situations, we extend the DBM to compressible two-component systems in a force field. Compared with the DBMs for single-component systems \cite{XuLai2016,XuChen2016}, this model is capable of obtaining more details of the flow field, such as density, hydrodynamic velocity, temperature, and pressure of each component. Moreover, this model has the merit of capturing the nonequilibrium manifestations of each component and the mixing entropy of the two-component flows. 

The rest of the paper is organized as follows. In Sec. \ref{SecII}, we propose the DBM which is a coarse-graining model of the Boltzmann equation. The collision term and two types of force terms are presented. Section \ref{SecIII} contains the numerical verification and validation of the DBM. Grid convergence tests are performed. In Sec. \ref{SecIV}, the nonequilibrium manifestations and entropy of mixing are investigated in the evolution of the RTI. The relevant physical phenomena and mechanisms are analyzed. Section \ref{SecV} gives conclusions and discussions.

\section{Discrete Boltzmann Equation}\label{SecII}

Recently, a DBM has been presented for a system containing two species, $\sigma=A$ and $B$ \cite{XuLin2016CNF}. The two species are described by two coupled discrete Boltzmann equations, which employ the same discrete velocity model. Hence, they have the same number of extra degrees of freedom $I^{\sigma}$ and specific-heat ratio $\gamma^{\sigma}$. Besides, no external force is under consideration in Ref. \cite{XuLin2016CNF}.

Here, we extend the DBM to a system containing two components with independent specific-heat ratios in a force field. The discrete Boltzmann equation takes the form,
\begin{equation}
\frac{\partial f^{\sigma}_{i}}{\partial t}+v^{\sigma}_{i\alpha}\frac{\partial f^{\sigma}_{i}}{\partial r_{\alpha}}=\Omega^{\sigma}_{i}+G^{\sigma}_{i}
\tt{,}
\label{DiscreteBoltzmannEquation}
\end{equation}
where ${f}^{\sigma eq}$ is the distribution function, $r_{\alpha}$ denotes the Cartesian coordinate in the $\alpha$ direction, $v^{\sigma}_{i\alpha}$ represents the discrete velocity, with $i=1$, $2$, $\dots$, $N$, and $N$ is the total number of discrete velocities. The collision term $\Omega^{\sigma}_{i}$ and force term $G^{\sigma}_{i}$ describe the rates of change in the distribution functions due to the molecular collision and external force, respectively.

\subsection{Collision term}

As is well known, it is difficult to solve the Boltzmann equation directly because of the complex collision term ${\Omega }^{\sigma }$. To make use of the Boltzmann equation, one needs to simplify the collision term. In the simplification (i.e., coarse-graining) process, the key is to keep the relevant physical quantities (such as density, velocity, and temperature) compatible with the Boltzmann equation \cite{Liu1990}. Generally, there are two steps in reducing the Boltzmann equation to the DBM.

\textbf{Step I: Linearization of collision term}

To be specific, the collision term is firstly linearized as follows,
\begin{equation}
{{\Omega }^{\sigma }}=-\frac{1}{{{\tau }^{\sigma }}}({{f}^{\sigma }}-{{f}^{\sigma eq}})  \tt{,}
\label{CollisionTerm1}
\end{equation}
where ${f}^{\sigma eq}$ is the local equilibrium distribution function; $\tau^{\sigma}=1/(n^{A}/\theta^{A}+n^{B}/\theta^{B})$ denotes the relaxation time dependent on the particle number density $n^{\sigma}$ and two flexible parameters ($\theta^{A}$, $\theta^{B}$) \cite{Sofonea2001}. Equation (\ref{CollisionTerm1}) satisfies the following condition
\begin{equation}
\int{\int{{{\Omega }^{\sigma }}\mathbf{\Psi }d\mathbf{v}d\eta }}=\int{\int{-\frac{1}{{{\tau }^{\sigma }}}({{f}^{\sigma }}-{{f}^{\sigma eq}})\mathbf{\Psi }d\mathbf{v}d\eta }}\tt{.}
\end{equation}
Here $\mathbf{\Psi}$ is a matrix whose elements are $\overbrace{\mathbf{vv}\cdots \mathbf{v}}^{\alpha }{{\eta }^{2\beta }}$, which is a $\alpha$-th order tensor, with $\alpha=0$, $1$, $\cdots$, and $\beta=0$, $1$, $\cdots$. In addition, $\mathbf{v}$ is the particle velocity describing translational motions, and $\eta^2$ is utilized to describe the internal energies in extra degrees of freedom corresponding to molecular rotation and/or vibration. It should be noted that the more elements the matrix $\mathbf{\Psi }$ has, the more complex the form of ${f}^{\sigma eq}$ is \cite{XuLin2016CNF}. Its form depends on the given elements, and it may be different in various modes, such as, the Bhatnagar-Gross-Krook (BGK) model \cite{BGK1954}, the Kogan model \cite{Kogan1958}, the ellipsoidal statistical BGK model \cite{Zhang2013}, the Shakhov model \cite{Shakhov1968}, the Rykov model \cite{Rykov1975}, the Liu model \cite{Liu1990}, etc. The expression of ${f}^{\sigma eq}$ adopted in this work reads \cite{Watari2007}
\begin{equation}
{{f}^{\sigma eq}}={{n}^{\sigma }}{{\left( \frac{{{m}^{\sigma }}}{2\pi kT} \right)}^{D/2}}{{\left( \frac{{{m}^{\sigma }}}{2\pi I^{\sigma} kT} \right)}^{1/2}}\exp \left[ -\frac{{{m}^{\sigma }}{{\left( \mathbf{v}-\mathbf{u} \right)}^{2}}}{2kT}-\frac{{{m}^{\sigma }}{{\eta }^{2}}}{2I^{\sigma} kT} \right]\tt{,}
\label{MostProbableDistribution}
\end{equation}
where $m^{\sigma}$ is the particle mass, $k=1$ the Boltzmann constant, $D=2$ the number of the spatial dimension, $\mathbf{u}$ the mixture velocity, $T$ the mixture temperature. Mathematically, Eq. (\ref{MostProbableDistribution}) is the most probable distribution in the system with given parameters ($m^{\sigma}$, $n^{\sigma}$, $\mathbf{u}$, $T$, $I^{\sigma}$), see Appendix \ref{APPENDIXA}. From Eq. (\ref{MostProbableDistribution}), we can obtain the moments of ${f}^{\sigma eq}$, see Eqs. (\ref{moment1})$-$(\ref{moment7}).

\textbf{Step II: Discretization of particle velocity}

The linearized Boltzmann equation is still difficult to solve as the term in Eq. (\ref{MostProbableDistribution}) depends on the particle velocity $\mathbf{v}$. The next step is to discretize the particle velocity, which aims to make the solution straightforward. Note that the quantities with which we are concerned are some kinetic moments that should remain unchanged in the simplification process \cite{Liu1990}. To achieve this aim, the moments calculated from the summation of ${f}^{\sigma eq}_{i}$ should be consistent with those from the integration of ${f}^{\sigma eq}$, where $f^{\sigma eq}_{i}$ is the discretization of ${f}^{\sigma eq}$ (similar for $f^{\sigma}_{i}$ and ${f}^{\sigma }$, ${{\Omega }^{\sigma }_{i}}$ and ${{\Omega }^{\sigma }}$).  Specifically,
\begin{equation}
\int{\int{{{f}^{\sigma eq}}}}d\mathbf{v}d\eta =\sum\nolimits_{i}{f_{i}^{\sigma eq}}
\tt{,}
\label{moment1}
\end{equation}
\begin{equation}
\int{\int{{{f}^{\sigma eq}}{{v}_{\alpha }}d\mathbf{v}d\eta }}=\sum\nolimits_{i}{f_{i}^{\sigma eq}{{v}^{\sigma}_{i\alpha }}}
\tt{,}
\label{moment2}
\end{equation}
\begin{equation}
\int{\int{{{f}^{\sigma eq}}}}\left( {{v}^{2}}+{{\eta }^{2}} \right)d\mathbf{v}d\eta =\sum\nolimits_{i}{f_{i}^{\sigma eq}\left( v_{i}^{\sigma 2}+\eta _{i}^{\sigma 2} \right)}
\tt{,}
\label{moment3}
\end{equation}
\begin{equation}
\int{\int{{{f}^{\sigma eq}}}}{{v}_{\alpha }}{{v}_{\beta }}d\mathbf{v}d\eta =\sum\nolimits_{i}{f_{i}^{\sigma eq}{{v}^{\sigma}_{i\alpha }}{{v}^{\sigma}_{i\beta }}}
\tt{,}
\label{moment4}
\end{equation}
\begin{equation}
\int{\int{{{f}^{\sigma eq}}\left( {{v}^{2}}+{{\eta }^{2}} \right){{v}_{\alpha }}d\mathbf{v}d\eta }}=\sum\nolimits_{i}{f_{i}^{\sigma eq}\left( v_{i}^{\sigma 2}+\eta _{i}^{\sigma 2} \right){{v}^{\sigma}_{i\alpha }}}
\tt{,}
\label{moment5}
\end{equation}
\begin{equation}
\int{\int{{{f}^{\sigma eq}}{{v}_{\alpha }}{{v}_{\beta }}{{v}_{\chi }}d\mathbf{v}d\eta }}=\sum\nolimits_{i}{f_{i}^{\sigma eq}{{v}^{\sigma}_{i\alpha }}{{v}^{\sigma}_{i\beta }}{{v}^{\sigma}_{i\chi }}}
\tt{,}
\label{moment6}
\end{equation}
\begin{equation}
\int{\int{{{f}^{\sigma eq}}\left( {{v}^{2}}+{{\eta }^{2}} \right){{v}_{\alpha }}{{v}_{\beta }}d\mathbf{v}d\eta }}
=\sum\nolimits_{i}{f_{i}^{\sigma eq}\left( v_{i}^{\sigma 2}+\eta _{i}^{\sigma 2} \right){{v}^{\sigma}_{i\alpha }}{{v}^{\sigma}_{i\beta }}}
\tt{,}
\label{moment7}
\end{equation}
where the integral is extended over the entire phase space ($\mathbf{v}$,$\eta $). The above seven equations are necessary for the recovery of NS equations with a flexible specific heat ratio in a force field. Note that the summations are exactly equal to the corresponding integrals. Substituting Eq. (\ref{MostProbableDistribution}) into Eqs. (\ref{moment1})-(\ref{moment7}) gives the results of kinetic moments. Then the discretization of Eq. (\ref{CollisionTerm1}) takes the form
\begin{equation}
{{\Omega }^{\sigma }_{i}}=-\frac{1}{{{\tau }^{\sigma }}}({{f}^{\sigma }_{i}}-{{f}^{\sigma eq}_{i}})
\label{CollisionTerm2}
\tt{,}
\end{equation}
where the discrete equilibrium distribution function $f^{\sigma eq}_{i}$ depends on the variables ($m^{\sigma}$, $n^{\sigma}$, $\mathbf{u}$, $T$, $I^{\sigma}$), i.e., $f^{\sigma eq}_{i}=f^{\sigma eq}_{i} (m^{\sigma}, n^{\sigma}, \mathbf{u}, T, I^{\sigma})$. Namely, the relationship between species is established in terms of the mixture velocity $\mathbf{u}$ and temperature $T$. Here the 2-dimensional 16-velocity model (D2V16) is adopted, see Ref.\cite{XuLin2016CNF} for more details. It should be mentioned that the species velocity $\mathbf{u}^{\sigma}$ may differ from the mixture velocity $\mathbf{u}$. The difference $\left(\mathbf{u}^{\sigma}-\mathbf{u}\right)$ refers to the diffusion velocity of species $\sigma$, and the term $\rho^{\sigma}\left(\mathbf{u}^{\sigma}-\mathbf{u}\right)$ is the diffusive flux of mass relative to the mixture velocity \cite{Sofonea2001}. Similarly, the species temperature $T^\sigma$ may be different from the mixture temperature $T$.

Remark: The collision term in Eq. (\ref{CollisionTerm2}) has two advantages over the previous one in Ref. \cite{XuLin2016CNF}. On the one hand, the extra degrees of freedom $I^{\sigma}$ in Eq. (\ref{MostProbableDistribution}) are independent for the two species. Consequently, the specific heat ratios $\gamma^{A}$ and $\gamma^{B}$ can be identical or different here. In contrast, they have the same values due to the constraint, $I^{A}=I^{B}$, in the previous work \cite{XuLin2016CNF}.
On the other hand, the two sets of parameters ($v^{A}_{i}$, $\eta^{A}_{i}$) and ($v^{B}_{i}$, $\eta^{B}_{i}$) can be either identical or different.
Thanks to this adjustability, the DBM could provide more versatile, robust, and accurate simulations. In contrast, the two sets of parameters are the same in Ref. \cite{XuLin2016CNF}, which is only a special case of this work.

\subsection{Force term}

In this subsection, we introduce two types of force terms. The first one, which was used in Ref. \cite{XuLai2016}, is extended to two-component systems here. The second is developed in this study. It is defined as the change of the discrete equilibrium distribution function over a short time interval due to the external force. 

\textbf{Type I}

Theoretically, the equilibrium distribution function is the leading part of the distribution function when the system is not too far from the equilibrium state \cite{XuLai2016}. In addition, $f^{\sigma}$ is closer to $f^{\sigma eq}( n^{\sigma}, \mathbf{u}^{\sigma}, T^{\sigma})$ than $f^{\sigma eq}( n^{\sigma}, \mathbf{u}, T)$. Consequently, the approximation $f^{\sigma }\approx f^{\sigma eq}( n^{\sigma}, \mathbf{u}^{\sigma}, T^{\sigma})$ is used to calculate the force term,
\begin{equation}
{{G}^{\sigma }}
=-\mathbf{a}\cdot \frac{\partial {f}^{\sigma}}{\partial \mathbf{v}}
\approx -\mathbf{a}\cdot \frac{\partial {f}^{\sigma eq}}{\partial \mathbf{v}}
=\frac{{{m}^{\sigma }}}{{{T}^{\sigma }}}\mathbf{a}\cdot \left( \mathbf{v}-{{\mathbf{u}}^{\sigma }} \right){{f}^{\sigma eq}}
\tt{,}
\end{equation}
whose discrete form is taken as
\begin{equation}
G_{i}^{\sigma }=\sum\nolimits_{\alpha}\frac{{{m}^{\sigma }}}{{{T}^{\sigma }}} {{a}_{\alpha }}\left( {{v}^{\sigma}_{i\alpha }}-u_{\alpha }^{\sigma } \right)f_{i}^{\sigma eq}
\tt{.}
\label{ForceTerm1}
\end{equation}
Here $\mathbf{a}=\sum_{\alpha}{a}_{\alpha }\mathbf{e}_{\alpha }$ represents the body acceleration and $\mathbf{e}_{\alpha }$ the unit vector in ${\alpha }$ direction.

\textbf{Type II}

In classical physics, the impulse (work) done by the external force changes the momentum (energy) of an object, and the mass remains constant in the force field. Similarly, it is regarded that the force has an influence on the velocity and energy of the fluid components, but it does not change the density. The equilibrium distribution function changes in the evolution of density, velocity, and temperature. Mathematically, the change of the equilibrium distribution function over a small time interval is an approximation of the change rate of the equilibrium distribution function. Consequently, the force term is obtained based on the assumption $f^{\sigma }\approx f^{\sigma eq}\left( n^{\sigma}, \mathbf{u}^{\sigma}, T^{\sigma}\right)$.

Specifically, due to the external force, the velocity and energy of component $\sigma$ change from $u_{\alpha }^{\sigma }$ and ${{E}^{\sigma }}$ into
\begin{equation}
u_{\alpha }^{\sigma \dagger }=u_{\alpha }^{\sigma }+{\tau }{{a}_{\alpha }}
\label{VelocityForce}
\tt{,}
\end{equation}
\begin{equation}
{{E}^{\sigma \dagger }}={{E}^{\sigma }}+\sum\nolimits_{\alpha}{\tau }{{\rho }^{\sigma }}u_{\alpha }^{\sigma }{{a}_{\alpha }}
\tt{,}
\label{EnergyForce}
\end{equation}
within the time interval ${\tau }$. Meanwhile, the temperature of component $\sigma$ becomes
\begin{equation}
{{T}^{\sigma \dagger }}={{T}^{\sigma }}-\frac{{m}^{\sigma}}{D+I^{\sigma}}{{\tau }^{2}}a_{\alpha }^{2}
\tt{,}
\end{equation}
which is derived from Eqs. (\ref{VelocityForce}), (\ref{EnergyForce}) and the definition ${{E}^{\sigma }}=\frac{D+I^{\sigma }}{2}{{n}^{\sigma }}{{T}^{\sigma }}+\frac{1}{2}{{\rho }^{\sigma }}{{u}^{\sigma 2}}$. Hence, the force term reads
\begin{equation}
G_{i}^{\sigma }=\frac{1}{{\tau }}\left[ f_{i}^{\sigma eq}\left( {{n}^{\sigma }},{{\mathbf{u}}^{\dagger \sigma }},{{T}^{\dagger \sigma }} \right)-f_{i}^{\sigma eq}\left( {{n}^{\sigma }},{{\mathbf{u}}^{\sigma }},{{T}^{\sigma }} \right) \right]
\tt{.}
\label{ForceTerm2}
\end{equation}

In addition, there are three similarities between the two types of force terms. (I) Based upon the approximation $f^{\sigma }\approx f^{\sigma eq}( n^{\sigma}, \mathbf{u}^{\sigma}, T^{\sigma})$, both of them are expressed with the discrete equilibrium distribution function; (II) Theoretically, the expressions in Eqs. (\ref{ForceTerm1}) and (\ref{ForceTerm2}) are equivalent at the level of the first order accuracy; (III) Both can be employed to recover the NS equations with force effects, see Appendix \ref{APPENDIXB}.

Obviously, the derivations of formulas (\ref{ForceTerm1}) and (\ref{ForceTerm2}) are different. Type I is conventional. Here we further explain Type II. In fact, it is straightforward to incorporate various effects (such as gravitation and chemical reaction) into the Boltzmann equation with the method of Type II in three steps: (i) we formulate the changes of physical quantities (e.g., density, velocity, and temperature) due to a given effect, and we calculate those quantities after a time step; (ii) we calculate the change of the discrete equilibrium distribution function, $\Delta f_{i}^{\sigma eq}$, within a time step, $\Delta t$; and (iii) the relevant effect term is obtained as, $\frac{\partial f_{i}^{\sigma }}{\partial t}|_{\text{effect}} \approx \frac{\Delta f_{i}^{\sigma eq}}{\Delta t}$, i.e., the change of the discrete equilibrium distribution function over a time step.

Finally, it is straightforward to extend the 2-dimensional (2D) DBM to a 3D DBM. The additional complexity for the 3D case is to calculate the discrete equilibrium distribution function, which should satisfy a required number of kinetic moments, i.e., ${\mathbf{\hat{f}}}^{eq}=\mathbf{M} {\mathbf{f}}^{eq}$. Here ${\mathbf{f}}^{eq}$ is a set of discrete equilibrium distribution functions, ${\mathbf{\hat{f}}}^{eq}$ is a set of kinetic moments, and $\mathbf{M}$ is a square matrix acting as a bridge between ${\mathbf{\hat{f}}}^{eq}$ and ${\mathbf{f}}^{eq}$. The discrete equilibrium distribution function takes the form ${\mathbf{f}}^{eq}=\mathbf{M}^{-1} {\mathbf{\hat{f}}}^{eq}$, which is the same as the 2D case  \cite{XuLin2016CNF}. The existence of $\mathbf{M}^{-1}$ should be ensured.

\section{Numerical validation}\label{SecIII}

In this section, we conduct numerical validation with two benchmark cases: binary diffusion and free fall. The former is to test the coupled collision terms with different discrete velocities and extra degrees of freedom, the latter is to validate the two types of force terms. We further compare the simulation results with the analytic solutions of the RTI in the next section. 

\subsection{Binary diffusion}

\begin{figure}[tbp]
	\begin{center}
		\includegraphics[bbllx=22pt,bblly=20pt,bburx=746pt,bbury=537pt,width=0.4\textwidth]{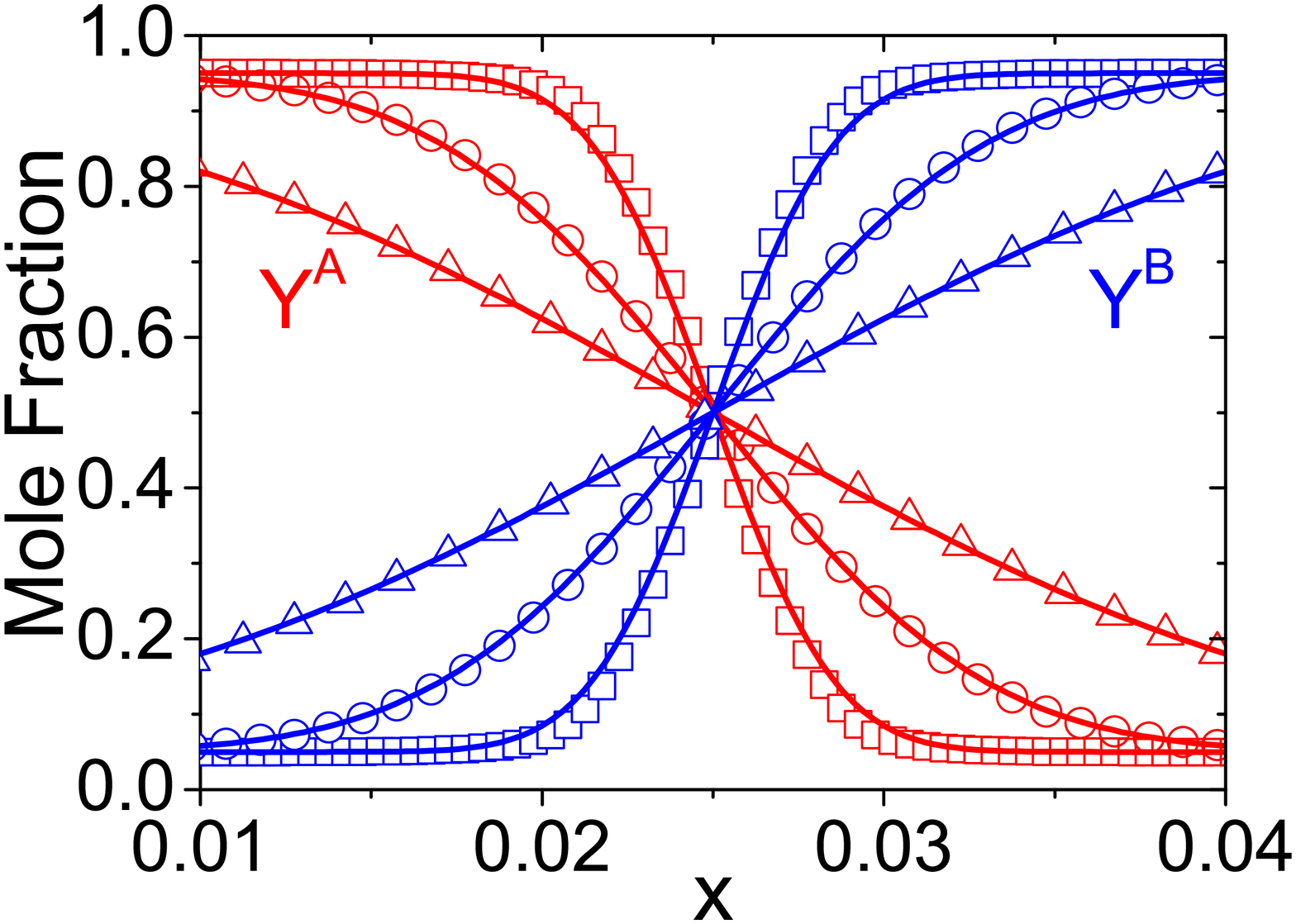}
	\end{center}
	\caption{Mole fractions ($Y^{A}$ and $Y^{B}$) versus $x$ in the binary diffusion. The squares, circles, and triangles denote simulation results at instants $t=0.004$, $0.02$, and $0.1$ respectively. The lines stand for corresponding analytic solutions.}
	\label{Fig01}
\end{figure}

There are two aims in simulating the binary diffusion. One is to validate the coupled collision terms $\Omega^{A}_{i}$ and $\Omega^{B}_{i}$ using different sets of discrete velocities. The other is to verify the independent specific heat ratios $\gamma^{A}$ and $\gamma^{B}$, which are functions of extra degrees of freedom $I^A$ and $I^B$, respectively. Here we set ($v^{A}_{a}$, $v^{A}_{b}$, $v^{A}_{c}$, $v^{A}_{d}$, $\eta^{A}_{a}$) $=$ ($0.2$, $0.9$, $1.1$, $2.5$, $2.9$), ($v^{B}_{a}$, $v^{B}_{b}$, $v^{B}_{c}$, $v^{B}_{d}$, $\eta^{B}_{a}$) $=$ ($0.3$, $0.9$, $1.1$, $2.6$, $3.9$), $I^A=3$ and $I^B=5$. A hydrostatic isothermal field is employed for the initial configuration, with the number densities, 
\begin{equation}
\left\{ \begin{matrix}
	{{\left( {{n}^{A}},{{n}^{B}} \right)}_{L}}=\left( 0.95,0.05 \right)  \\
	{{\left( {{n}^{A}},{{n}^{B}} \right)}_{R}}=\left( 0.05,0.95 \right)  \\
\end{matrix} \right.
\nonumber
\tt{,}
\end{equation}
where the suffix $L$ indexes the left part $x<0.025$, $R$ the right part $x \ge 0.025$. The time step is $\Delta t=2 \times 10^{-5}$, the space step $\Delta x=\Delta y=5 \times 10^{-4}$, the grid $N_x \times N_y=100 \times 1$. Moreover, the inflow and outflow boundary conditions are employed in the horizontal direction, and the periodic boundary conditions are applied in the vertical direction. 

Figure \ref{Fig01} illustrates the mole fractions $Y^A$ and $Y^B$ along $x$ at times $t=0.004$, $0.02$ and $0.1$, respectively. The symbols denote DBM results. The lines represent analytic solutions, $Y^{\sigma}=\frac{1}{2}+\frac{\Delta Y^{\sigma}}{2}$ erf($\frac{x}{\sqrt{4 D t}}$), with the diffusion coefficient $D=p \tau \rho (\rho^A \rho^B)^{-1} Y^{A}Y^{B}$ and the initial mole fraction difference $\Delta Y^{\sigma}=0.9$. The simulation results agree well with the analytic solutions. Note that it is more accurate and robust to set $\eta^{A}_{a}>\eta^{B}_{a}$ for $I^A>I^B$, as demonstrated by a series of simulations (not shown here). And it is preferable to choose the discrete velocities ($v^{\sigma}_{a}$, $v^{\sigma}_{b}$, $v^{\sigma}_{c}$, $v^{\sigma}_{d}$) around the hydrodynamic velocity and sound speed.

Moreover, it is worth mentioning that the implementation of the DBM simulation is quite simple based on the linear Eq. (\ref{DiscreteBoltzmannEquation}). Here we adopt the forward Euler scheme (with first order accuracy) for time discretization, and the nonoscillatory and nonfree-parameters dissipative finite difference scheme (with second order accuracy) for space discretization \cite{Zhang1991NND}. In fact, computational costs of the proposed model are modest. It takes only $7$ seconds to complete the simulation in Fig. \ref{Fig01}, using a computational facility with Intel(R) Core(TM) i7-6700K CPU @ 4.00GHz and RAM 32.00 GB.

\subsection{Free fall}

\begin{figure}[tbp]
	\begin{center}
		\includegraphics[bbllx=77pt,bblly=62pt,bburx=543pt,bbury=277pt,width=0.7\textwidth]{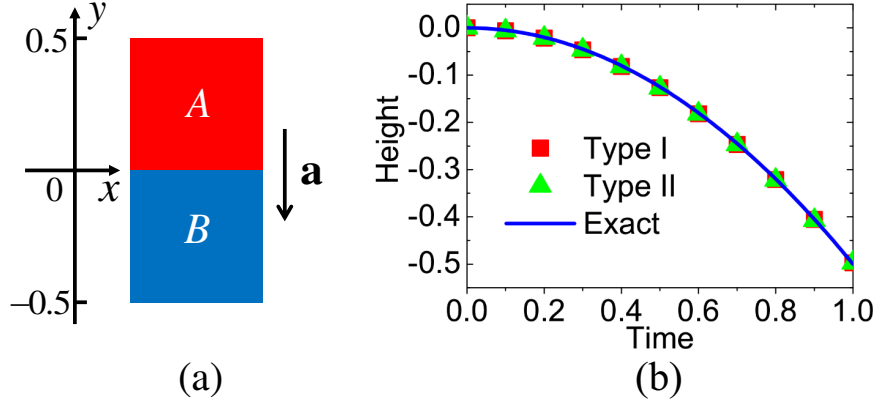}
	\end{center}
	\caption{Initial configuration for the free fall case (a) and the heigh of material interface (b): Type I (squares), Type II (triangles), and the exact solution (line).}
	\label{Fig02}
\end{figure}

For the sake of testing the force effect, the motion of free fall is simulated. The two types of force terms in Eqs. (\ref{ForceTerm1}) and (\ref{ForceTerm2}) are adopted, respectively. The simulation is performed on a grid $N_{x}\times N_{y}=1\times 400$, with a space step $\Delta x=\Delta y=5\times 10^{-4}$ and a time step $\Delta t=10^{-5}$. The computational domain is initially divided into two regions, $-0.5 \le y \le 0$ and $0 < y \le 0.5$. The upper half is filled with medium $A$ and the lower with $B$. The pressure $p=1$ is uniform in the whole domain. The temperature in the upper (lower) part is $T_{u}=1$ ($T_{d}=2$). The particle number density is $p/T_{u}$ for species $A$, and $p/T_{d}$ for $B$. Other parameters are $a_{x}=0$, $a_{y}=-1$, $m^{\sigma}=1$, $\theta^{\sigma}=1.5\times 10^{-5}$, $I^{\sigma}=3$, ($v^{\sigma}_{a}$, $v^{\sigma}_{b}$, $v^{\sigma}_{c}$, $v^{\sigma}_{d}$, $\eta^{\sigma}_{a}$) $=$ ($0.6$, $1.1$, $2.1$, $2.9$, $1.9$). Moreover, the periodic boundary conditions are employed in the horizontal direction, and the inflow and outflow boundary conditions are specified in the vertical direction. 

Figure \ref{Fig02} delineates the initial configuration for the free fall, and it shows the height of the material interface between the two media. The simulation results from formulas (\ref{ForceTerm1}) and (\ref{ForceTerm2}), respectively, are compared with the exact solution, $y=0.5 a_{y} t^{2}$. 
Initially, the material interface is located at height $y=0$. It falls down with acceleration ($0$, $a_y$) as time goes on. 
Clearly, all the results are in perfect agreement. It is demonstrated that the DBM is suitable for a system in a force field, and both types of force terms work well. 

\section{RTI}\label{SecIV}

The DBM for single component systems can only be used to simulate the RTI in a special situation with superposition of a heavy cold medium above a hot light one \cite{XuLai2016,XuChen2016}. Our improved DBM for two-component systems has the capability of investigating the RTI in more complex cases where the two media have independent temperatures. In this work, the case of a uniform initial temperature field is investigated, which is beyond the capability of DBMs for single-component systems \cite{XuLai2016,XuChen2016}. Moreover, compared with the DBMs for single-component systems \cite{XuLai2016,XuChen2016}, this model can be used to probe more details of the system, such as the density, hydrodynamic velocity, temperature, and pressure of each component.

This section consists of three subsections. In the first part, we construct two initial configurations for the RTI in two-component compressible flows. The grid convergence analysis is performed, and a comparison is made between the numerical and analytic results. In the next subsection, the invariants of tensors for nonequilibrium quantities are proposed and investigated in the dynamic processes of RTI. Finally, we probe the entropy of mixing in the evolution of the RTI.

\subsection{Flow Field}

With the gravitational acceleration $\mathbf{a}=(0,-g)$, the initial flow field is on the hydrostatic unstable equilibrium, i.e.,
\begin{equation}
\nabla p=\rho \mathbf{a}
\tt{,}
\label{nablapressure}
\end{equation}
which means increasing pressure from top to bottom. To satisfy this condition, two initial configurations are proposed as below,
\begin{eqnarray}
\left\{
\begin{array}{l}
T={{T}_{u}}, {{n}^{A}}=\frac{{{p}_{m}}}{{{T}_{u}}}\exp \left[ \frac{{{m}^{A}}g}{{{T}_{u}}}\left( y_{m}-{y} \right) \right], {{n}^{B}}=0, y>{{y}_{m}} \tt{,} \\
T={{T}_{d}}, {{n}^{B}}=\frac{{{p}_{m}}}{{{T}_{d}}}\exp \left[ \frac{{{m}^{B}}g}{{{T}_{d}}}\left( y_{m}-{y} \right) \right], {{n}^{A}}=0, y<{{y}_{m}} \tt{,}
\end{array}
\right.
\label{configurationA}
\end{eqnarray}
or
\begin{eqnarray}
\left\{
\begin{array}{l}
{{n}^{A}}={{n}_{u}}, {{n}^{B}}=0, T={{m}^{A}}g\left( y_{m}-{y} \right)+\frac{{{p}_{m}}}{{{n}_{u}}}, y>{{y}_{m}} \tt{,} \\
{{n}^{B}}={{n}_{d}}, {{n}^{A}}=0, T={{m}^{B}}g\left( y_{m}-{y} \right)+\frac{{{p}_{m}}}{{{n}_{d}}}, y<{{y}_{m}} \tt{,}
\end{array}
\right.
\label{configurationB}
\end{eqnarray}
where the subscripts $u$ and $d$ denote the upper and lower parts of the physical domain, $m$ represents the material interface. Considering the transition layer across the material interface, the field jump is smoothed by a tanh profile with width $W$. Specifically, the initial temperature profile is chosen to be
$T=(T_u+T_d)/2+(T_u-T_d)/2\times \tanh((y-y_m)/W)$
in Eq. (\ref{configurationA}), and the initial densities are
$n^A=n_u/2+n_u/2\times \tanh((y-y_m)/W)$ and $n^B=n_d/2-n_d/2\times \tanh((y-y_m)/W)$ in Eq. (\ref{configurationB}). A half single-mode sinusoidal perturbation with amplitude $A_{0}$ is imposed on the interface, i.e., $y_m=L_y/2 + A_{0} \cos(\pi x/L_x)$. Here $L_x$ and $L_y$ are the width and length of the system, respectively. Furthermore, the symmetrical boundary conditions are adopted in the $x$ direction, and the specular reflection boundary conditions are imposed in the $y$ direction. 

\begin{center}
	\begin{table}[tbp]
		\begin{tabular}{cccccc}
			\hline\hline
			Cases & $Re$ & $\theta^{\sigma}$ & $\Delta t$ & $\Delta x=\Delta y$ & $
			\begin{array}{c}
			\left(
			v^{A}_{a}, v^{A}_{b}, v^{A}_{c}, v^{A}_{d}, \eta^{A}_{a}
			\right) \\
			\left(
			v^{B}_{a}, v^{B}_{b}, v^{B}_{c}, v^{B}_{d}, \eta^{B}_{a}
			\right)
			\end{array}
			$ \\
			\hline
			Run I & $2000$ & $1.262\times10^{-4}$ & $5\times10^{-6}$ & $1\times 10^{-4}$ & $
			\begin{array}{c}
			\left(
			5.5, 2.5, 0.7, 0.9, 5.3
			\right) \\
			\left(
			6.0, 2.7, 0.2, 0.5, 6.3
			\right)
			\end{array}
			$ \\
			\hline
			Run II & $500$ & $5.046\times10^{-4}$ & $2\times10^{-5}$ & $2\times 10^{-4}$ & $
			\begin{array}{c}
			\left(
			4.5, 2.2, 0.2, 0.5, 5.3
			\right) \\
			\left(
			6.0, 2.7, 0.3, 0.9, 6.3
			\right)
			\end{array}
			$ \\
			\hline
			Run III & $125$ & $2.018\times10^{-3}$ & $4\times10^{-5}$ & $4\times 10^{-4}$ & $
			\begin{array}{c}
			\left(
			4.5, 2.2, 0.2, 0.5, 5.3
			\right) \\
			\left(
			6.0, 2.7, 0.3, 0.9, 6.3
			\right)
			\end{array}
			$ \\
			\hline\hline
		\end{tabular}
		\caption{Parameters for the RTI with various Reynolds numbers.}
		\label{TableI}
	\end{table}
\end{center}

Let us consider an initial isothermal configuration described by Eq. (\ref{configurationA}) \footnote{Other initial configurations can also be adopted, such as an isentropic initial configuration or a constant-density configuration described by Eq. (\ref{configurationB}). In this work, the physical conclusions are insensitive to the initial configurations.}. Three runs are carried out with Reynolds numbers $Re=2000$, $500$, and $125$, respectively. Mathematically, the Reynolds number is a function of space and time in the evolution of a compressible flow. Here we define $Re=\bar{\rho }\bar{u}\bar{L}/\bar{\mu }$, where the characteristic length is $\bar{L}=\lambda$, characteristic velocity $\bar{u}=\sqrt{g/k}$, and the wave number $k=2\pi/\lambda$. The characteristic density $\bar{\rho }=\sum\nolimits_{\sigma }{{{m}^{\sigma }}n_{m}^{\sigma }M_{m}^{\sigma }}$, the dynamic viscosity $\bar{\mu }={{p}_{m}}{{\tau }_{m}}$, the relaxation time ${{\tau }_{m}}={{\left( \sum\nolimits_{\sigma }{n_{m}^{\sigma }Y_{m}^{\sigma }{{\theta }^{\sigma -1}}} \right)}^{-1}}$, and the particle number density $n_{m}^{\sigma }$, mole fraction $Y_{m}^{\sigma }$, and pressure $p_{m}^{\sigma }$ are initially located at the material interface. The parameters for the three runs are $m^{A}=3$, $m^{B}=1$, $T_u=T_d=1.0$, $W=L_y/200$, $A_{0}=L_y/100$, $p_{m}=40$, $g=2$, $I^{\sigma}=3$, and $L_{x}\times L_{y}=0.1\times 1.0$; the other parameters are listed in Table \ref{TableI}. Additionally, the reduced time $t^*=t(g/\lambda)^{1/2}$, reduced length $L^*=L/\lambda$, and reduced velocity $u^*=u(g\lambda)^{-1/2}$ are employed in this work. 

\begin{figure}[tbp]
	\begin{center}
		\includegraphics[bbllx=2pt,bblly=36pt,bburx=557pt,bbury=214pt,width=0.9\textwidth]{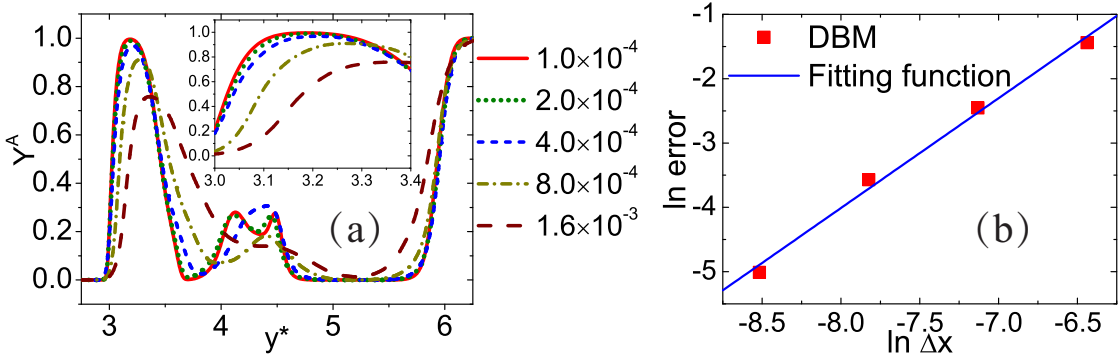}
	\end{center}
	\caption{Comparison among the simulation results with various space steps. (a) Mole fraction $Y^A$ versus height $y^*$ along the grid $x^*=0.64$ at time $t^*=3.79$ in the evolution of RTI with $Re=2000$. The simulation is conducted with five different space steps: $\Delta x=\Delta y=1.0\times 10^{-4}$ (solid), $2.0\times 10^{-4}$ (dot), $4.0\times 10^{-4}$ (short dash), $8.0\times 10^{-4}$ (dash dot), and $1.6\times 10^{-3}$ (long dash). (b) The relation between space steps and numerical errors. The squares stand for the DBM results and the line represents the fitting function.}
	\label{Fig03}
\end{figure}

First of all, to verify the resolution, a grid convergence analysis is performed. As an example, the RTI with $Re=2000$ is simulated using five different space steps $\Delta x=\Delta y=1.0\times 10^{-4}$, $2.0\times 10^{-4}$, $4.0\times 10^{-4}$, $8.0\times 10^{-4}$, and $1.6\times 10^{-3}$, respectively. The other parameters are the same as those for Run I in Table \ref{TableI}. Figure \ref{Fig03} (a) shows the mole fraction, $Y^A=n^{A}/(n^{A}+n^{B})$, versus height $y^*$ along the grid $x^*=0.75$ at time $t^*=3.79$. The insert map in Fig \ref{Fig03} (a) is the enlargement of the horizontal scale $3.0 \leq x \leq 3.4$. It can be found that the simulation results are converging with decreasing (increasing) space steps (resolution). In particular, the results with space steps $1.0\times 10^{-4}$ and $2.0\times 10^{-4}$ are quite close to each other. Consequently, the resolution ($1.0\times 10^{-4}$) is satisfactory for Run I.

For the sake of a quantitative study, Fig. \ref{Fig03} (b) illustrates the relation between space steps and numerical errors. The errors refer to the height differences of the leftmost peak with $\Delta x =\Delta y=1.0 \times 10^{-4}$ and the leftmost peak in the other four cases. The squares stand for the DBM results, and the line represents the fitting function, $\ln(\mathrm{error}) =  1.7 \ln (\Delta x) + 9.6$. It is demonstrated that the numerical error reduces with decreasing space step. In a similar way, the grid convergence analysis can be studied with Run II or III in Table \ref{TableI}, whose resolution is high enough as well. 

\begin{figure}[tbp]
	\begin{center}
		\includegraphics[bbllx=115pt,bblly=50pt,bburx=579pt,bbury=248pt,width=1.\textwidth]{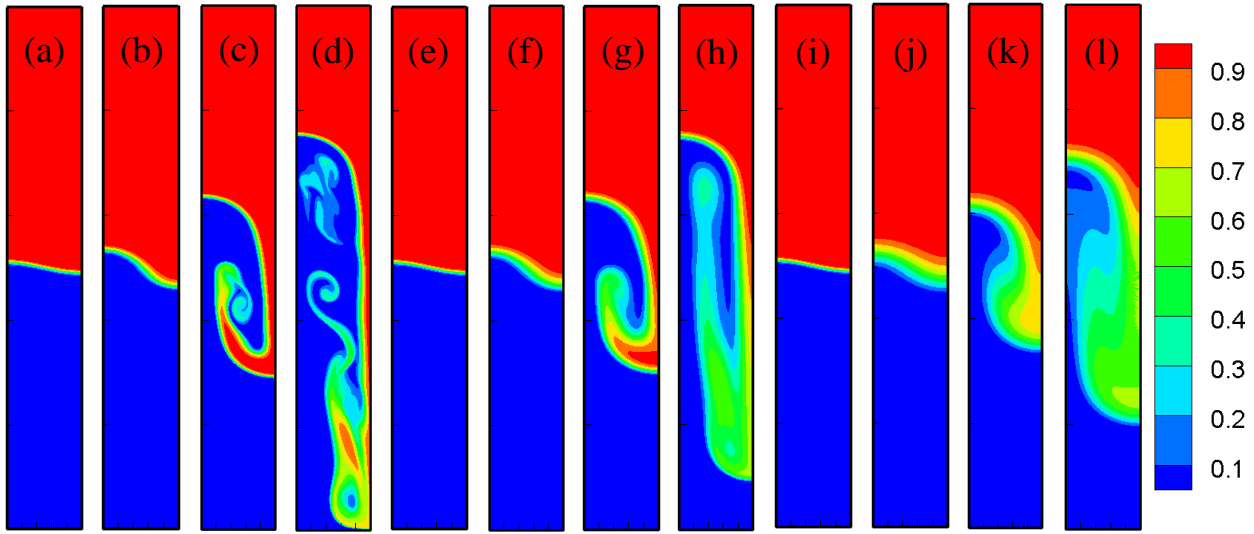}
	\end{center}
	\caption{Contours of mole fraction $Y^A$ in the evolution of RTI with various Reynolds numbers at times $t^*=0.0$, $1.26$, $3.79$, and $6.32$, respectively. (a)-(d) are for $Re=2000$, (e)-(h) are for $Re=500$, (i)-(l) are for $Re=125$.}
	\label{Fig04}
\end{figure}

Figure \ref{Fig04} depicts the contours of mole fraction $Y^A$ in the evolution of the RTI. Subplots (a)-(d) are for the case $Re=2000$, (e)-(h) are for $Re=500$, (i)-(l) are for $Re=125$. In each case, we display the snapshots at times $t^*=0.0$, $1.26$, $3.79$, and $6.32$, respectively. (I) Initially, the movements of the two media are symmetrical in all cases. Later, the material interface gets farther and farther away from the sinusoidal shape. The heavy fluid drops down and forms a spike with time, while the light fluid rises up with the formation of a bubble. In addition, the two media penetrate into each other as time advances. (II) With opposite movements of the two species, the Kelvin-Helmholtz instability  (KHI) \cite{Kelvin1871,Helmholtz1868,Wang2009,XuGan2011} takes place when the tangential velocity varies across the material interface. Owing to the KHI, the spike rolls up and the mushroom structure emerges, which promotes the fluid mixing process. (III) With the increase of dynamic viscosity (i.e., the decrease of $Re$), the width of the material interface increases fast, the KHI is suppressed, and the downward-moving spike becomes slow. The simulation results show similar behaviors to those in Refs. \cite{Wei2012,Liang2014}. 

\begin{figure}[tbp]
	\begin{center}
		\includegraphics[bbllx=21pt,bblly=82pt,bburx=553pt,bbury=266pt,width=0.8\textwidth]{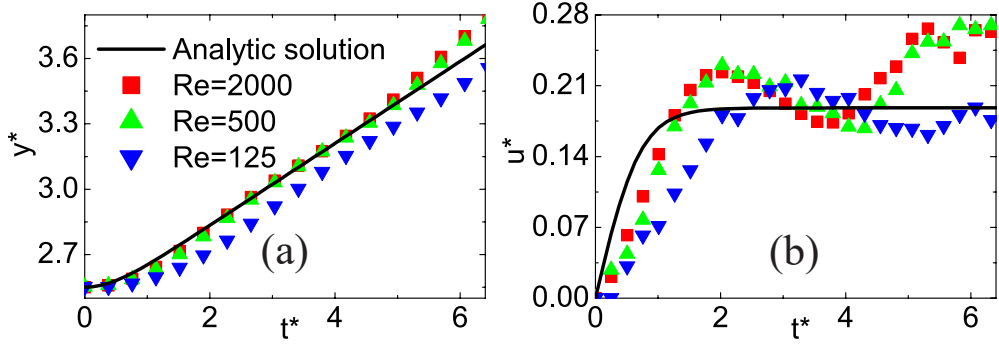}
	\end{center}
	\caption{The height $y^{*}$ (a) and speed $u^{*}$ (b) of the bubble front versus time $t^{*}$. The solid lines denote the analytic solutions, the symbols denote simulation results with various Reynolds numbers: $Re=2000$ (squares), $500$ (upper triangles), and $125$ (lower triangles).}
	\label{Fig05}
\end{figure}

Let us illustrate the evolution of the bubble front in Fig. \ref{Fig05}. Panels (a) and (b) show its height $y^*$ and speed $u^*$ versus time $t^*$, respectively. The squares, upper triangles, and lower triangles stand for the DBM results with $Re=2000$, $500$, and $125$, respectively. The solid lines represent the analytic solutions calculated by Eqs. (10) and (11) in Ref. \cite{Tao2012}. It can be found in Figs. \ref{Fig05} (a) and (b) that the simulation results are roughly around the analytic solutions. As shown in Fig. \ref{Fig05} (b), there are four distinctive stages in the evolution of the bubble speed, i.e., exponential growth, potential flow growth, reacceleration, and chaotic development \cite{Wei2012}. For example, in the case $Re=2000$, the four stages are $0 \le t^*<1.90$, $1.90 \le t^*<3.67$, $3.67\le t^*<5.31$, and $t^* \ge 5.31$, respectively. In the exponential growth stage (described by the linear stability theory), the small amplitude perturbation grows exponentially with time. Afterwards, the potential flow growth (described by the potential flow theory) stage is characterized by a quasiconstant bubble front speed, which was often mistaken for ``terminal velocity" \cite{Wei2012}. During the reacceleration stage, the bubble is reaccelerated due to the rotating vortices with the development of KHI. The bubble speed fluctuates with time, and the flow reaches the turbulent mixing state in the chaotic development stage. Four such regimes were also obtained in previous works \cite{Wei2012,Liang2014}. In contrast, the analytic solutions demonstrate only two simple stages, namely the acceleration and plateau stages. There are six main reasons for the differences between the present simulation results and the analytic solutions in Ref. \cite{Tao2012}. 

(I) The DBM is for viscid flow, whereas the theory is for inviscid flow. Physically, the flow viscosity widens the material interface and impedes the growth of the RTI. One can find in panels (a) and (b) that the simulation results with higher Reynolds numbers (i.e., lower viscosities) are closer to the analytic solutions where the viscosity is completely ignored  \cite{Tao2012}.

(II) The DBM is for compressible flow, while the theory is for incompressible flow. From top to bottom of both half parts, the initial density and pressure increase gradually in our simulations, while they remain constants in the theory \cite{Tao2012}. Meanwhile, the Atwood number, which enhances the RTI, changes with space and time in our simulations, whereas it remains constant in the simple theory \cite{Tao2012}.

(III) The DBM works also for the flow with local rotations, while the theory considers only the irrotational case. The vortices induced by the KHI result in the oscillations of the $u^*$. Specifically, the bubble speed is sensitive to the bypass flow around the vortex. The flow velocity behind the upper surface of the bubble oscillates as the vortices emerge, expand, split, merge, and break up, etc. From the DBM results, we can also observe that the viscosity suppresses the creation and development of the vortex, and it delays the oscillation of $u^*$.

(IV) The energy transformation in the DBM is different from that in the theory. The gravitational potential energy converts into the kinetic, compressive, and internal energies in the DBM simulation, which is more consistent with the real case \cite{XuLai2016}, whereas it only changes into the kinetic energy in the consideration of theory. Moreover, the vortex splits the kinetic energy, in the DBM results and the real case, so that the part moving upward decreases, while the kinetic energy is not split in the theory. This is one reason why the bubble speed from our simulation is slower than that from the theory in the initial stage (for $t^*=0$ to $1.2$).

(V) As for our initial configuration, the density varies continuously across the interface which is more consistent with the real situation. But in the theory, the interface is just assumed as a strong discontinuity. Consequently, the local Atwood number in our DBM simulation is much smaller than that in the theory. This is another reason why the bubble speed in the DBM is slower than in the theory in the early period.

(VI) The boundary conditions for our DBM and for the theory are different in the $y$ direction. The specular reflection boundary condition is used for the DBM, while the infinite boundary condition is used for the theory \cite{Tao2012}. Consequently, the bubble is bounded within the box in the simulation, while its height always increases in the theory.

In summary, the DBM results are more reasonable than the analytic solutions which do not consider the viscosity, compressibility, or rotation at all \cite{Tao2012}. 

\subsection{Nonequilibrium manifestations}

In reality, almost all physical systems evolve in nonequilibrium states. The HNE and TNE behaviors have essential influences upon nonequilibrium systems. Generally, the nonequilibrium state is too complex to be described. Fortunately, the DBM provides an effective tool to investigate the nonequilibrium manifestations \cite{XuLai2016,XuChen2016,XuLin2014PRE,XuLin2015PRE,XuGan2015SM,XuLin2016CNF}. It is noteworthy that $f^{\sigma eq}$ (and $f^{\sigma eq}_{i}$) can be replaced by $f^{\sigma}$ (and $f^{\sigma}_{i}$) in Eqs. (\ref{moment1})-(\ref{moment3}) according to the conservation of mass, momentum and energy. Such a replacement may break Eqs. (\ref{moment4})-(\ref{moment7}), since high order kinetic moments of $f^{\sigma}$ (and $f^{\sigma}_{i}$) differ from those of $f^{\sigma eq}$ (and $f^{\sigma eq}_{i}$) in the nonequilibrium state. In fact, the nonequilibrium effects can be measured by the following terms
\begin{equation}
\mathbf{\Delta }_{2}^{\sigma *}=\sum\nolimits_{i}{{{m}^{\sigma }}\left( f_{i}^{\sigma }-f_{i}^{\sigma eq} \right)\mathbf{v}_{i}^{\sigma *}\mathbf{v}_{i}^{\sigma*}}
\label{Delta2} \tt{,}
\end{equation}
\begin{equation}
\mathbf{\Delta }_{3,1}^{\sigma *}=\sum\nolimits_{i}{{{m}^{\sigma }}\left( f_{i}^{\sigma }-f_{i}^{\sigma eq} \right)\left( v_{i}^{\sigma *2}+\eta _{i}^{\sigma 2} \right)\mathbf{v}_{i}^{\sigma *}}
\label{Delta31} \tt{,}
\end{equation}
\begin{equation}
\mathbf{\Delta }_{3}^{\sigma *}=\sum\nolimits_{i}{{{m}^{\sigma }}\left( f_{i}^{\sigma }-f_{i}^{\sigma eq} \right)\mathbf{v}_{i}^{\sigma *}\mathbf{v}_{i}^{\sigma *}\mathbf{v}_{i}^{\sigma *}}
\label{Delta3} \tt{,}
\end{equation}
\begin{equation}
\mathbf{\Delta }_{4,2}^{\sigma *}=\sum\nolimits_{i}{{{m}^{\sigma }}\left( f_{i}^{\sigma }-f_{i}^{\sigma eq} \right)\left( v_{i}^{\sigma *2}+\eta _{i}^{\sigma 2} \right)\mathbf{v}_{i}^{\sigma *}\mathbf{v}_{i}^{\sigma *}}
\label{Delta42} \tt{,}
\end{equation}
where $\mathbf{v}^{\sigma *}_{i}=\mathbf{v}^{\sigma}_{i}-\mathbf{u}$ is the particle velocity $\mathbf{v}^{\sigma}_{i}$ relative to the hydrodynamic velocity $\mathbf{u}$.

The components of the above tensors in Eqs. (\ref{Delta2})-(\ref{Delta42}) depend on the coordinate system. They may have different results if the coordinate system changes. For the sake of convenience, we introduce the invariants of tensors that are coefficients of the characteristic polynomial of the tensors. The invariants do not change with rotation of the coordinate system (they are objective), and they have significant physical meanings. Let us write a tensor as $\mathbf{\Delta }$ and its invariant as $\Lambda$. The invariant of the first-order tensor whose components are ${\Delta }_{\alpha}$ is
\begin{equation}
\Lambda=\sum\nolimits_{\alpha}{{{\Delta }_{\alpha}}{{\Delta }_{\alpha}}}
\tt{.}
\label{InvariantI}
\end{equation}
The independent invariants of the second-order tensor having components ${\Delta }_{\alpha \beta}$ are
\begin{equation}
\Lambda_{1}=\sum\nolimits_{\alpha}{{{\Delta }_{\alpha \alpha}}}
\tt{,}
\end{equation}
\begin{equation}
\Lambda_{2}=\sum\nolimits_{\alpha, \beta}{{{\Delta }_{\alpha \beta}}{{\Delta }_{\alpha \beta}}}
\label{InvariantII}
\tt{.}
\end{equation}
The independent invariants of the third-order tensor with components ${\Delta }_{\alpha \beta \gamma }$ are
\begin{equation}
\Lambda_{1}=\sum\nolimits_{\alpha ,\beta ,\gamma }{{{\Delta }_{\alpha \beta \gamma }}{{\Delta }_{\alpha \beta \gamma }}}
\tt{,}
\end{equation}
\begin{equation}
\Lambda_{2}=\sum\nolimits_{\alpha ,\beta ,\gamma }{{{\Delta }_{\alpha \alpha \gamma }}{{\Delta }_{\beta \beta \gamma }}}
\tt{,}
\end{equation}
\begin{equation}
\Lambda_{3}=\sum\nolimits_{\alpha ,\beta ,\gamma ,\delta ,\varepsilon ,\zeta }{{{\Delta }_{\alpha \beta \gamma }}{{\Delta }_{\gamma \delta \varepsilon }}{{\Delta }_{\delta \varepsilon \zeta }}{{\Delta }_{\alpha \beta \zeta }}}
\tt{.}
\label{InvariantIII3}
\end{equation}
The subscripts $\alpha ,\beta ,\gamma ,\delta ,\varepsilon ,\zeta$ in Eqs. (\ref{InvariantI})-(\ref{InvariantIII3}) stand for $x$ or $y$. To be specific, the invariants of the second-order tensor $\mathbf{\Delta }_{2}^{\sigma *}$ are $\Lambda^{\sigma *}_{1}$ and $\Lambda^{\sigma *}_{2}$. $\frac{1}{2}\Lambda^{\sigma *}_{1}$ denotes the ``fluctuation of translational energy". Physically, $\Lambda^{\sigma *}_{2}$ is defined as the value of ``non-organized stress", and mathematically, $\Lambda^{\sigma *}_{2} \geq 0$. The invariant $\Lambda^{\sigma *}$ of vector $\mathbf{\Delta }_{3,1}^{\sigma *}$ is twice the value of ``non-organized heat flux". The term ``non-organised" is relative to ``organised". Organised behaviors refer to the collective motion of the fluid, while non-organised behaviors correspond to the molecular individualism on top of the collective motion \cite{XuLai2016}.

\begin{figure}[tbp]
	\begin{center}
		\includegraphics[bbllx=112pt,bblly=349pt,bburx=397pt,bbury=753pt,width=0.66\textwidth]{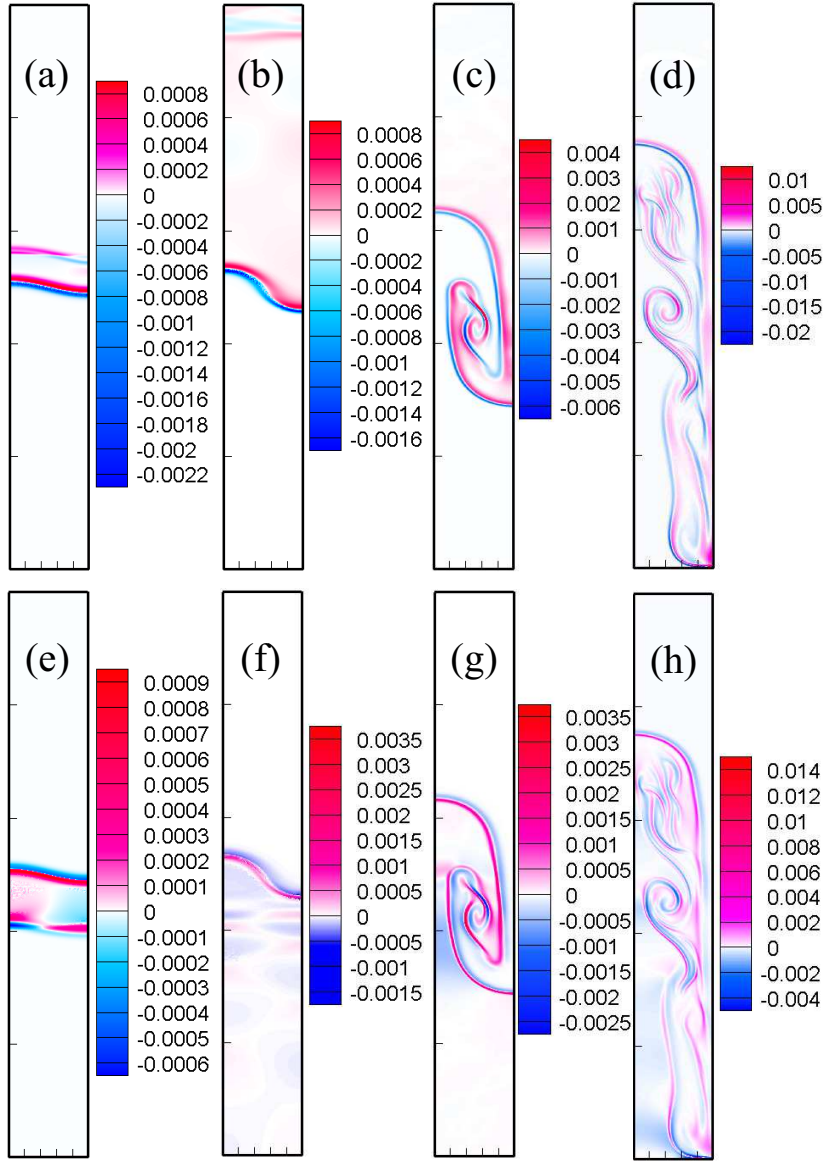}
	\end{center}
	\caption{Contours of invariant $\Lambda^{\sigma *}_{1}$ of the tensor $\mathbf{\Delta }_{2}^{\sigma *}$ in the case of $Re=2000$ at times $t^*=0.126$, $1.26$, $3.79$, and $6.32$, from left to right, respectively. The first row (a)-(d) is for $\Lambda^{A *}_{1}$ and the second row (e)-(h) for $\Lambda^{B *}_{1}$.}
	\label{Fig06}
\end{figure}

Firstly, we investigate the invariant $\Lambda^{\sigma *}_{1}$ of the tensor $\mathbf{\Delta }_{2}^{\sigma *}$. Figure \ref{Fig06} illustrates the contours of $\Lambda^{A *}_{1}$ (a)-(d) and $\Lambda^{B *}_{1}$ (e)-(h) in the case of $Re=2000$ at times $t^*=0.126$, $1.26$, $3.79$, and $6.32$, respectively. Because the initial fields are not naturally continuous across the interface, two perturbation waves emerge and leave the initial interface in opposite directions at sound speed. Nonequilibrium effects take place around the material interface and perturbation waves where physical gradients exist, see Figs. 5 (a), (b), (e) and (f). Then, they diminish and vanish eventually with the dissipation of perturbation waves, see Figs. 5 (c), (d), (g) and (h).

It can be found in Fig. \ref{Fig06} that, across the interface, the topological structures of $\Lambda^{A *}_{1}$ and $\Lambda^{B *}_{1}$ are quite similar and their values are qualitatively opposite, because the physical gradients of species $A$ and $B$ are opposite. The two invariants are (non-)zero where the system is in the (non-)equilibrium state. The value of $\Lambda^{\sigma *}_{1}$ being positive (or negative) means the translational energy $\frac{1}{2}\sum\nolimits_{\alpha}{\sum\nolimits_{i}{{{m}^{\sigma }}f_{i}^{\sigma}{v}_{i \alpha}^{\sigma * 2}}}$ is greater (or less) than its equilibrium counterpart $\frac{1}{2}\sum\nolimits_{\alpha}{\sum\nolimits_{i}{{{m}^{\sigma }}f_{i}^{\sigma eq}{v}_{i \alpha}^{\sigma * 2}}}$ where the energy equipartition theorem breaks. Around the material interface where the physical gradients are sharp, the system deviates far from the equilibrium state, and the invariants show a substantial departure from zero. Moreover, the areas for $\Lambda^{\sigma *}_{1}\neq 0$ increase as the material interface is enlarged in the evolution of the RTI. $\Lambda^{\sigma *}_{1}$ increases dramatically when the interface reaches the bottom, and the physical gradients become significantly sharper.

\begin{figure}[tbp]
	\begin{center}
		\includegraphics[bbllx=5pt,bblly=209pt,bburx=589pt,bbury=790pt,width=0.99\textwidth]{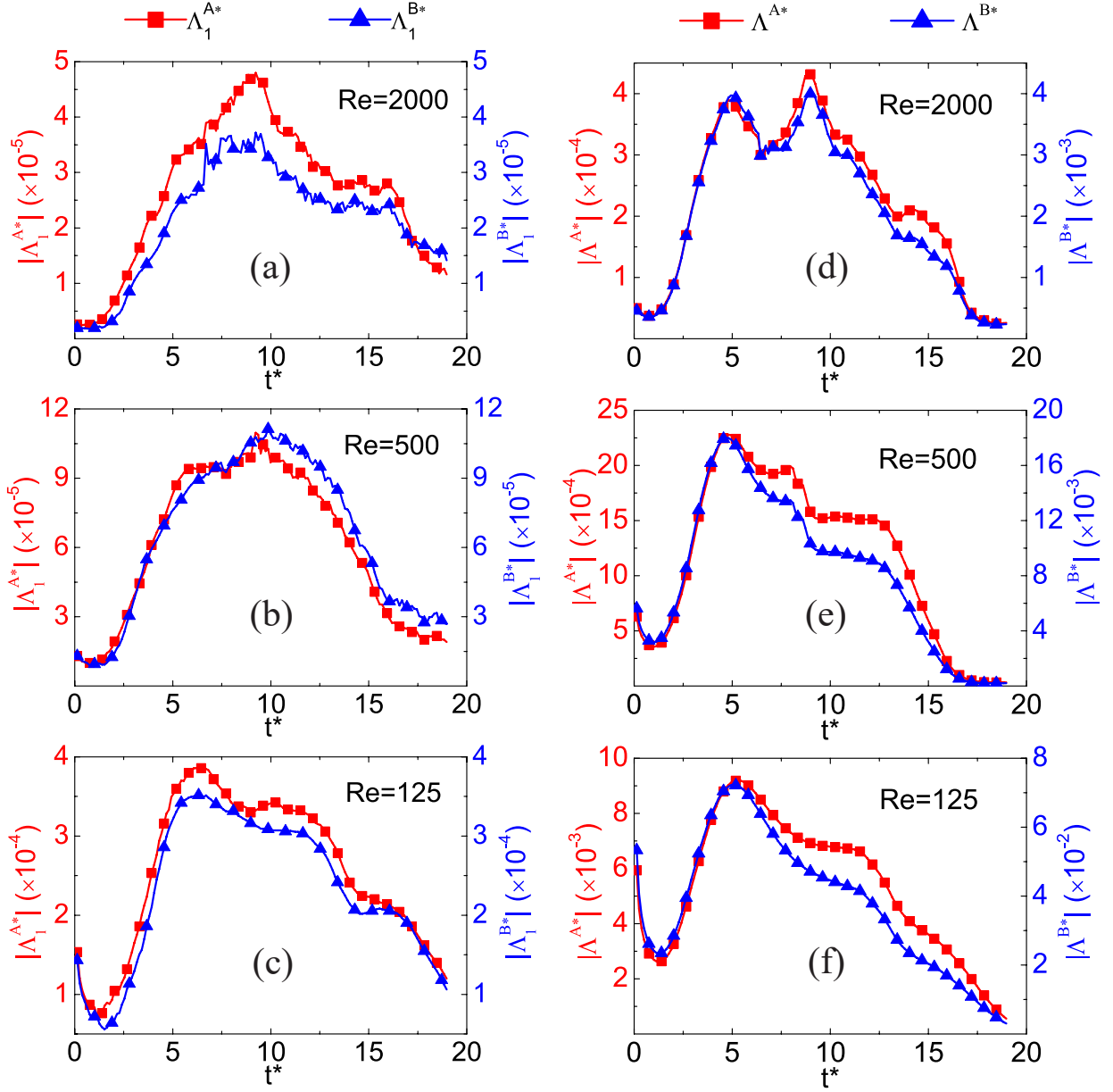}
	\end{center}
	\caption{The evolution of $\int\int{|\Lambda^{\sigma *}_{1}| dxdy}$ and $\int\int{|\Lambda^{\sigma *}| dxdy}$. Here $\Lambda^{\sigma *}_{1}$ is the invariant of the tensor $\mathbf{\Delta }_{2}^{\sigma *}$, and $\Lambda^{\sigma *}$ the invariant of the tensor $\mathbf{\Delta }_{3,1}^{\sigma *}$. Panels (a)-(c) are for $\int\int{|\Lambda^{\sigma *}_{1}| dxdy}$ with $Re=2000$, $500$, and $125$, respectively. Panels (d)-(f) are for $\int\int{|\Lambda^{\sigma *}| dxdy}$ with $Re=2000$, $500$, and $125$, respectively. Species $A$ and $B$ are denoted by lines with squares and triangles, respectively.}
	\label{Fig07}
\end{figure}

Let us probe the global TNE intensity from the viewpoints of the ``fluctuation of translational energies" ($\frac{1}{2}\Lambda^{\sigma *}_{1}$) and ``non-organized heat flux" ($\frac{1}{2}\Lambda^{\sigma *}$), respectively. As mentioned above, $\Lambda^{\sigma *}_{1}$ is the invariant of the tensor $\mathbf{\Delta }_{2}^{\sigma *}$ in Eq. (\ref{Delta2}), and $\Lambda^{\sigma *}$ is the invariant of the tensor $\mathbf{\Delta }_{3,1}^{\sigma *}$ in Eq. (\ref{Delta31}). Figure \ref{Fig07} illustrates the evolution of $\int\int{|\Lambda^{\sigma *}_{1}| dxdy}$ and $\int\int{|\Lambda^{\sigma *}| dxdy}$, where the integral extends over the whole computational domain. Panels (a)-(c) are for $\int\int{|\Lambda^{\sigma *}_{1}| dxdy}$ with $Re=2000$, $500$, and $125$, respectively. Panels (d)-(f) are for $\int\int{|\Lambda^{\sigma *}| dxdy}$ with $Re=2000$, $500$, and $125$, respectively. Species $A$ and $B$ are denoted by lines with squares and triangles, respectively. The left axes and ticks are for $A$, and the right for $B$. The orders of magnitude of those quantities are shown
after the axis titles. It is easy to find the similarities and differences between the variable tendencies in panels (a)-(f).

First of all, the values of $\int\int{|\Lambda^{\sigma *}_{1}| dxdy}$ and $\int\int{|\Lambda^{\sigma *}| dxdy}$ demonstrate similar trends. For the low Reynolds number, there are three obviously different stages, i.e., the early decreasing, intermediate increasing, and final decreasing tendencies. For the high Reynolds number, the nonequilibrium effects are weak, and the early reducing trend is particularly suppressed and even eliminated. Hence, with the increasing Reynolds number, the three tendencies may be reduced to two (increasing and decreasing tendencies) in the whole process, see panels (a) and (d). For all cases, the nonequilibrium manifestations approach zero when the mixing between media reaches saturation eventually. 

In addition, there is a smooth transition between the two regimes with low and high Reynolds numbers. From $Re=125$ to $500$, and then to $2000$, both $\int\int{|\Lambda^{\sigma *}_{1}| dxdy}$ and $\int\int{|\Lambda^{\sigma *}| dxdy}$ reduce gradually. Namely, TNE has weaker global influences on the system with a higher Reynolds number. Moreover, as shown in panels (a)-(c), the values of $\int\int{|\Lambda^{A *}_{1}| dxdy}$ and $\int\int{|\Lambda^{B *}_{1}| dxdy}$ are roughly close to each other, while $\int\int{|\Lambda^{A *}| dxdy}$ is smaller than $\int\int{|\Lambda^{B *}| dxdy}$ by approximately an order of magnitude in panels (d)-(f). For example, $\int\int{|\Lambda^{A *}_{1}| dxdy}=9.1 \times 10^{-3}$ and $\int\int{|\Lambda^{B *}_{1}| dxdy}=7.2 \times 10^{-2}$ at time $t^*=5$ in panel (f). The physical reason for the considerable difference between $\int\int{|\Lambda^{A *}| dxdy}$ and $\int\int{|\Lambda^{B *}| dxdy}$ is that $\int\int{|\Lambda^{\sigma *}| dxdy}$ is inversely proportional to $m^{\sigma 2}$ and $m^{A}:m^{B}=3:1$.

In fact, there are competitive effects on the nonequilibrium manifestations in the process of the RTI. Mathematically, the global TNE intensity increases with increasing relaxation time, nonequilibrium area, and/or values of physical gradients \cite{XuLin2016CNF}. Physically, for a long relaxation time, the system departs far from the local nonequilibrium, and the transport coefficient (such as the dynamic viscosity coefficient or heat conductivity) is large. Meanwhile, a fast transportation process smoothes the physical gradient quickly, hence the nonequilibrium effects diminish fast. In fact, the nonequilibrium manifestations are intense for sharp physical gradients \cite{XuLin2016CNF}. Hence, the nonequilibrium effects are weak when the material interface is wide. As time goes on, the nonequilibrium area enlarges, and the physical gradients become smoothed. Consequently, the values of $\int\int{|\Lambda^{\sigma *}_{1}| dxdy}$ show different tendencies in the whole process.

It is worth mentioning that the invariants ($\Lambda^{\sigma *}_{1}$, $\Lambda^{\sigma *}_{2}$, $\Lambda^{\sigma *}_{3}$) of the tensor $\mathbf{\Delta }_{3}^{\sigma *}$ have behaviors quite similar to those in Fig. \ref{Fig07} (not shown here). Furthermore, the nonequilibrium quantities are not susceptible to numerical errors in the discretization and evolution of the DBM \cite{XuLin2015PRE,XuLin2016CNF}. It can be verified that those quantities are physical but not artificial results of DBM. For example, the results of $\int\int{|\Lambda^{A *}| dxdy}$ and $\int\int{|\Lambda^{B *}| dxdy}$ are $4.88 \times 10^{-4}$ and $3.73 \times 10^{-3}$ for $Re=2000$ at $t^*=3.79$ in Fig. \ref{Fig07} (d) with the space step $\Delta x=\Delta y=10^{-4}$, whereas they are $5.07 \times 10^{-4}$ and $4.20 \times 10^{-3}$ with $\Delta x=\Delta y=2\times 10^{-4}$. The relative differences are $4\%$ and $13\%$ , respectively. This is satisfactory.

\subsection{Entropy of mixing}

Entropy is of great concern and interest to both physicists and engineers. In thermodynamics, the entropy of mixing is a portion of the total entropy. It increases when several separate parts of the components are mixed without chemical reaction before the establishment of a new thermodynamic equilibrium state in a closed system. Compared with DBMs for single-component systems \cite{XuLai2016,XuChen2016}, this DBM has the capability of investigating entropy of mixing,
\begin{equation}
S_M=-\sum\nolimits_{\sigma }{{{n}^{\sigma }}\ln {{Y}^{\sigma }}}
\tt{.}
\end{equation}
In this subsection, we investigate the entropy of mixing in the evolution of the RTI.

\begin{figure}[tbp]
	\begin{center}
		\includegraphics[bbllx=111pt,bblly=554pt,bburx=354pt,bbury=752pt,width=0.5\textwidth]{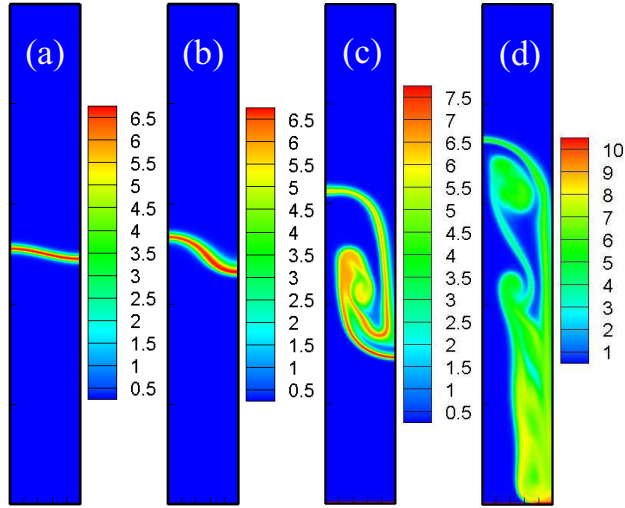}
	\end{center}
	\caption{Snapshots of entropy of mixing in the evolution of RTI with $Re=2000$ at times $t^*=0.126$, $1.26$, $3.79$, and $6.32$, from left to right, respectively.}
	\label{Fig08}
\end{figure}
\begin{figure}[tbp]
	\begin{center}
		\includegraphics[bbllx=8pt,bblly=402pt,bburx=442pt,bbury=556pt,width=0.8\textwidth]{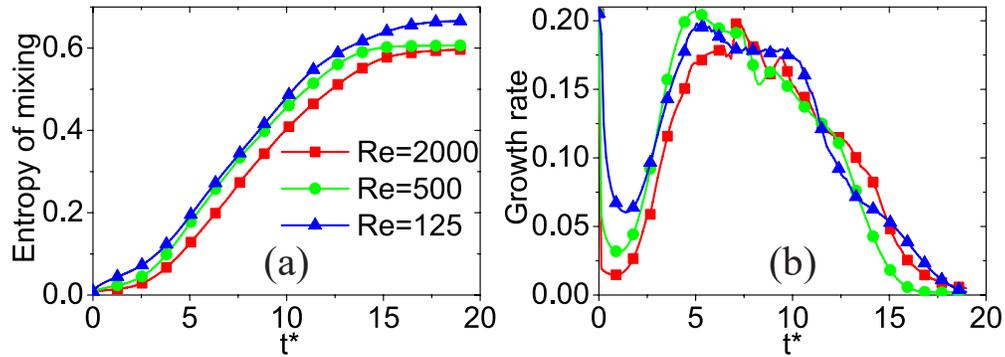}
	\end{center}
	\caption{The integral of the entropy of mixing (a) and its growth rate (b) in the evolution of the RTI with various Reynolds numbers: $Re=2000$, $500$, and $125$, respectively.}
	\label{Fig09}
\end{figure}

Figure \ref{Fig08} exhibits the snapshots of entropy of mixing in the process of the RTI with $Re=2000$ at times $t^*=0.126$, $1.26$, $3.79$, and $6.32$ from left to right, respectively. Obviously, the entropy of mixing is zero in the single-component area, and it is greater than zero around the material interface between the two miscible media. The area for $S_M>0$ enlarges in the mixing process as the material interface widens and lengthens.

To have a quantitative study, we calculate the integral of the entropy of mixing, $\int\int{S dxdy}$, where the integral is extended over the whole region. Figure \ref{Fig09} depicts the integral (a) and its growth rate (b) in the evolution of the RTI with $Re=2000$, $500$, and $125$, respectively. 
On the whole, the entropy of mixing increases monotonously and the growth rate shows three stages, i.e., the initial decreasing, intermediate increasing, and eventual decreasing trends. 
Initially, the growth rate (i.e., diffusion speed) reduces as the density gradient becomes smooth. The mixing speed is high for the large diffusion coefficient $D$ that is proportional to the relaxation time $\tau$. 
Later, the growth rate increases with the increasing length of the material interface. 
Finally, it becomes slow again when the mixing starts to saturate. The entropy of mixing tends towards the maximum and its growth rate approaches zero as the media are mixed adequately. 

In fact, the material interface enlarges and the density gradients become smooth in the process of the RTI. The increasing material interface and the smoothing density gradients have opposite influences on the mixing speed. On the one hand, the smoothing density gradients impede the mixing. On the other hand, the increasing material interface promotes the mixing. Consequently, the mixing speed reduces (increases) when the smoothing density gradient (increasing material interface) dominates. In fact, the growth rate of the entropy of mixing and the global TNEs have similar competitive mechanisms. 

\section{Conclusions and discussions}\label{SecV}

A discrete Boltzmann model (DBM) is proposed for two-component compressible flows. Previous single-component DBMs are suitable for the RTI only in a special situation with superposition of a heavy cold medium above a hot light one \cite{XuLai2016,XuChen2016}, whereas this two-component DBM is capable of more practical simulations where the two media can have independent temperatures. This DBM also has great advantages over another DBM for two-component flows \cite{XuLin2016CNF}. Specifically, the discrete velocity model is the same in the two discrete Boltzmann equations, the specific-heat ratios are identical for different species, and no external force is considered in the previous model \cite{XuLin2016CNF}. The present DBM is suitable for more general cases in which discrete velocity models are independent in the discrete Boltzmann equations, the specific-heat ratios are flexible for two components, and the external force is taken into account. Moreover, two types of force terms are exploited to describe the gravitational effects. In addition to the recovery of the modified NS equations in the hydrodynamic limit, the DBM has the capability of measuring nonequilibrium behaviors in various complex fluid systems. Numerical validation and verification are performed via two benchmark cases, i.e., binary diffusion and free fall. 

We use the DBM to investigate the RTI. The numerical results are compared with the analytic solutions. Their differences are analyzed from six viewpoints: (i) viscosity, (ii) compressibility, (iii) local rotation, (iv) energy transformation, (v) initial configuration, and (vi) boundary condition. Moreover, we introduce and study the invariants of tensors for nonequilibrium effects. Specifically, we probe the global TNE effects, including the ``fluctuation of translational energies" and ``non-organized heat flux". For the low Reynolds number, they first decrease, then increase and finally decrease. For the high Reynolds number, the early reducing trend is suppressed and even eliminated. It is interesting to find that the growth rate of entropy of mixing demonstrates similar tendencies to the global TNE effects. Because there are similar competitive effects on the nonequilibrium manifestations and the growth rate of entropy of mixing: (i) the increasing material interface enhances them; (ii) the reducing physical gradients weaken them; (iii) the long relaxation time enlarges them; (iv) the large viscosity impedes the growth of material interface. 

\begin{acknowledgments}
	
The authors thank Drs. Huilin Lai, Yanbiao Gan, Zhipeng Liu, Yudong Zhang, and Linlin Fei for fruitful discussions. AX and GZ acknowledge support of National Natural Science Foundation of China [under Grant Nos. 11475028 and 11772064], Science Challenge Project (under Grant No. JCKY2016212A501). KL and CL acknowledge support from the Center for Combustion Energy at Tsinghua University. YL and CL acknowledge support of the National Natural Science Foundation of China [under Grant Nos. 11574390, 41472130, and 11374360], National Basic Research Program of China [under Grant No. 2013CBA01504].
CL acknowledges support of the China Postdoctoral Science Foundation (under Grant No. 2017M620757), and FJKLMAA, Fujian Normal University.

\end{acknowledgments}

\appendix

\section{}\label{APPENDIXA}

Here we provide the derivation of the equilibrium distribution function in Eq. (\ref{MostProbableDistribution}). For this sake, we introduce the entropy,
\begin{equation}
S^{\sigma }=-k\left\langle {{f}^{\sigma }}\ln {{f}^{\sigma }} \right\rangle
\label{entropy_defination}
\tt{,}
\end{equation}
with the operator $\left\langle {} \right\rangle =\iint{d\mathbf{v}d\eta }$. Mathematically, the maximum of entropy is the function of the most probable distribution. Physically, the entropy has the maximum when the system reaches the equilibrium state. Consequently, the most probable distribution is regarded as the equilibrium distribution function. For a 2D system, Eq. (\ref{entropy_defination}) can be rewritten as
\begin{align}
	& S^{\sigma }=-k\left\langle {{f}^{\sigma }}\ln {{f}^{\sigma }} \right\rangle +{{L}_{1}}\left( \left\langle {{f}^{\sigma }} \right\rangle -\hat{f}_{1}^{\sigma } \right)+{{L}_{2}}\left( \left\langle {{f}^{\sigma }}{{v}_{x}} \right\rangle -\hat{f}_{2}^{\sigma } \right)+{{L}_{3}}\left( \left\langle {{f}^{\sigma }}{{v}_{y}} \right\rangle -\hat{f}_{3}^{\sigma } \right) \nonumber \\
	&  +{{L}_{4}}\left( \left\langle {{f}^{\sigma }}{{v}^{2}} \right\rangle -\hat{f}_{4}^{\sigma } \right)+{{L}_{5}}\left( \left\langle {{f}^{\sigma }}{{\eta }^{2}}/{{I}^{\sigma }} \right\rangle -\hat{f}_{5}^{\sigma } \right)
	\tt{,}
\end{align}
with $\hat{f}_{1}^{\sigma }=\left\langle {{f}^{\sigma }} \right\rangle$, $\hat{f}_{2}^{\sigma }=\left\langle {{f}^{\sigma }}{{v}_{x}} \right\rangle$, $\hat{f}_{3}^{\sigma }=\left\langle {{f}^{\sigma }}{{v}_{y}} \right\rangle$, $\hat{f}_{4}^{\sigma }=\left\langle {{f}^{\sigma }}{{v}^{2}} \right\rangle$, and $\hat{f}_{5}^{\sigma }=\left\langle {{f}^{\sigma }}{{\eta }^{2}}/{{I}^{\sigma }} \right\rangle$. A slight variation of the distribution function $\delta f ^{\sigma}$ results in a small change of entropy,
\begin{align}
	& \delta S^{\sigma }=-k\left\langle \ln {{f}^{\sigma }}\delta {{f}^{\sigma }} \right\rangle -k\left\langle \delta {{f}^{\sigma }} \right\rangle +{{L}_{1}}\left( \left\langle \delta {{f}^{\sigma }} \right\rangle -\delta \hat{f}_{1}^{\sigma } \right)+{{L}_{2}}\left( \left\langle \delta {{f}^{\sigma }}{{v}_{x}} \right\rangle -\delta \hat{f}_{2}^{\sigma } \right)  \nonumber \\
	&  +{{L}_{3}}\left( \left\langle \delta {{f}^{\sigma }}{{v}_{y}} \right\rangle -\delta \hat{f}_{3}^{\sigma } \right)+{{L}_{4}}\left( \left\langle \delta {{f}^{\sigma }}{{v}^{2}} \right\rangle -\delta \hat{f}_{4}^{\sigma } \right)+{{L}_{5}}\left( \left\langle \delta {{f}^{\sigma }}{{\eta }^{2}}/{{I}^{\sigma }} \right\rangle -\delta \hat{f}_{5}^{\sigma } \right)
	\label{deltaS}
	\tt{.}
\end{align}
In thermodynamic equilibrium (${{f}^{\sigma }}={{f}^{\sigma eq}}$), the system has maximum entropy, i.e.,
\begin{equation}
{{\left. \frac{\delta S^{\sigma }}{\delta {{f}^{\sigma }}} \right|}_{\delta {{f}^{\sigma }}\to 0}}\left( {{f}^{\sigma eq}} \right)=0
\label{maximum}
\tt{.}
\end{equation}
Substituting Eq. (\ref{deltaS}) into (\ref{maximum}), we can obtain
\begin{equation}
0=-k\ln {{f}^{\sigma eq}}-k+{{L}_{1}}+{{L}_{2}}{{v}_{x}}+{{L}_{3}}{{v}_{y}}+{{L}_{4}}{{v}^{2}}+{{L}_{5}}{{\eta }^{2}}/{{I}^{\sigma }}
\tt{,}
\end{equation}
i.e.,
\begin{equation}
{{f}^{\sigma eq}}=\exp \left( -1+L_{1}^{*}+L_{2}^{*}{{v}_{x}}+L_{3}^{*}{{v}_{y}}+L_{4}^{*}{{v}^{2}}+L_{5}^{*}{{\eta }^{2}}/{{I}^{\sigma }} \right)
\label{f_L}
\tt{,}
\end{equation}
with $L_{j}^{*}\text{=}{{L}_{j}}/k$, $j=1$, $\dots$, $5$.
For ${{f}^{\sigma }}={{f}^{\sigma eq}}$, the following relations
\begin{equation}
\left\{ \begin{array}{*{35}{l}}
	\left\langle {{f}^{\sigma }} \right\rangle ={{n}^{\sigma }}=\hat{f}_{1}^{\sigma }  \\
	\left\langle {{f}^{\sigma }}{{v}_{x}} \right\rangle ={{n}^{\sigma }}{{u}_{x}}=\hat{f}_{2}^{\sigma }  \\
	\left\langle {{f}^{\sigma }}{{v}_{y}} \right\rangle ={{n}^{\sigma }}{{u}_{y}}=\hat{f}_{3}^{\sigma }  \\
	\left\langle {{f}^{\sigma }}{{v}^{2}} \right\rangle ={{n}^{\sigma }}\left( DkT/{{m}^{\sigma }}+{{u}^{2}} \right)=\hat{f}_{4}^{\sigma }  \\
	\left\langle {{f}^{\sigma }}{{\eta }^{2}}/{{I}^{\sigma }} \right\rangle ={{n}^{\sigma }}kT/{{m}^{\sigma }}=\hat{f}_{5}^{\sigma }  \\
\end{array} \right.
\label{relations}
\tt{,}
\end{equation}
are satisfied according to the conservation laws and the energy equipartition theorem.
From Eqs. (\ref{f_L}) and (\ref{relations}), we can obtain
\begin{equation}
\left\langle {{f}^{\sigma }} \right\rangle ={{n}^{\sigma }}=\hat{f}_{1}^{\sigma }={{\left( -L_{4}^{*} \right)}^{-1}}{{\left( -L_{5}^{*}/{{I}^{\sigma }} \right)}^{-1/2}}{{\pi }^{3/2}}\exp \left( -1+L_{1}^{*}-\frac{L_{2}^{*2}+L_{3}^{*2}}{4L_{4}^{*}} \right)
\label{average_1}
\tt{,}
\end{equation}
and
\begin{equation}
{{u}_{x}}={{\bar{v}}_{x}}=\frac{\left\langle {{f}^{\sigma }}{{v}_{x}} \right\rangle }{\left\langle {{f}^{\sigma }} \right\rangle }=\frac{\left\langle {{\partial }_{L_{\text{2}}^{\text{*}}}}{{f}^{\sigma }} \right\rangle }{\left\langle {{f}^{\sigma }} \right\rangle }=\frac{{{\partial }_{L_{\text{2}}^{\text{*}}}}\hat{f}_{\text{1}}^{\sigma }}{\hat{f}_{\text{1}}^{\sigma }}=-\frac{L_{\text{2}}^{\text{*}}}{2L_{\text{4}}^{\text{*}}}
\tt{,}
\end{equation}
where the operator $\bar{\psi }$ denotes the average value of $\psi $ in the phase space ($\mathbf{v}$,$\eta $). In a similar way, we can obtain
\begin{equation}
{{u}_{y}}=\frac{{{\partial }_{L_{3}^{*}}}\hat{f}_{1}^{\sigma }}{\left\langle {{f}^{\sigma }} \right\rangle }=-\frac{L_{3}^{*}}{2L_{4}^{*}}
\tt{,}
\end{equation}
\begin{equation}
DkT/{{m}^{\sigma }}+{{u}^{2}}=\frac{{{\partial }_{{{L}_{4}}}}\hat{f}_{1}^{\sigma }}{\left\langle {{f}^{\sigma }} \right\rangle }=\frac{L_{2}^{*2}+L_{3}^{*2}-4L_{4}^{*}}{4L_{4}^{*2}}
\tt{,}
\end{equation}
\begin{equation}
T=\frac{{{\partial }_{L_{5}^{*}}}\hat{f}_{1}^{\sigma }}{\left\langle {{f}^{\sigma }} \right\rangle }=-\frac{1}{2L_{5}^{*}}
\label{average_5}
\tt{.}
\end{equation}
From Eqs. (\ref{average_1})-(\ref{average_5}), we obtain
\begin{equation}
\left\{ \begin{array}{*{35}{l}}
	L_{1}^{*}=\ln {{n}^{\sigma }}-\ln \left[ {{\left( \frac{{{m}^{\sigma }}}{2kT} \right)}^{-1}}{{\left( \frac{{{m}^{\sigma }}}{2{{I}^{\sigma }}kT} \right)}^{-1/2}}{{\pi }^{3/2}} \right]+1-\frac{{{u}^{2}}}{2T}  \\
	L_{2}^{*}=\frac{{{m}^{\sigma }}{{u}_{x}}}{kT}  \\
	L_{3}^{*}=\frac{{{m}^{\sigma }}{{u}_{y}}}{kT}  \\
	L_{4}^{*}=-\frac{{{m}^{\sigma }}}{2kT}  \\
	L_{5}^{*}=-\frac{{{m}^{\sigma }}}{2kT}  \\
\end{array} \right.
\label{Eq_L}
\tt{.}
\end{equation}
Substituting Eq. (\ref{Eq_L}) into Eq. (\ref{f_L}) gives
\begin{equation}
{{f}^{\sigma eq}}={{n}^{\sigma }}\left( \frac{{{m}^{\sigma }}}{2\pi kT} \right){{\left( \frac{{{m}^{\sigma }}}{2\pi {{I}^{\sigma }}kT} \right)}^{1/2}}\exp \left[ -\frac{{{m}^{\sigma }}{{\left( \mathbf{v}-\mathbf{u} \right)}^{2}}}{2kT}-\frac{{{m}^{\sigma }}{{\eta }^{2}}}{2{{I}^{\sigma }}kT} \right]
\tt{.}
\end{equation}
The 3D equilibrium distribution function can be obtained in a similar way. It can be written in a uniform form in Eq. (\ref{MostProbableDistribution}).

\section{}\label{APPENDIXB}

It is easy to prove that, via the Chapman-Enskog multiscale analysis, this model could recover the NS equations in a force field in the hydrodynamic limit \cite{XuLin2016CNF,XuLai2016},
\begin{equation}
\label{NS_sigma_1}
\frac{\partial {{\rho }^{\sigma }}}{\partial t}+\frac{\partial }{\partial {{r}_{\alpha }}}\left( {{\rho }^{\sigma }}u_{\alpha }^{\sigma } \right)=0
\tt{,}
\end{equation}%
\begin{eqnarray}
\label{NS_sigma_2}
& \frac{\partial }{\partial t}\left( {{\rho }^{\sigma }}u_{\alpha }^{\sigma } \right)+\frac{\partial }{\partial {{r}_{\beta }}}\left( {{\delta }_{\alpha \beta }}{{p}^{\sigma }}+{{\rho }^{\sigma }}u_{\alpha }^{\sigma }u_{\beta }^{\sigma } \right)+\frac{\partial }{\partial {{r}_{\beta }}}\left( P_{\alpha \beta }^{\sigma }+U_{\alpha \beta }^{\sigma } \right) \nonumber \\
& ={{\rho }^{\sigma }}{{a}_{\alpha }}-\frac{{{\rho }^{\sigma }}}{{{\tau }^{\sigma }}}\left( u_{\alpha }^{\sigma }-{{u}_{\alpha }} \right)
\tt{,}
\end{eqnarray}%
\begin{eqnarray}
\label{NS_sigma_3}
& \frac{\partial }{\partial t}{{\rho }^{\sigma }}\left( {{e}^{\sigma }}+\frac{1}{2}{{u}^{\sigma 2}} \right)+\frac{\partial }{\partial {{r}_{\alpha }}}\left[ {{\rho }^{\sigma }}u_{\alpha }^{\sigma }\left( {{e}^{\sigma }}+\frac{1}{2}{{u}^{\sigma 2}} \right)+{{p}^{\sigma }}u_{\alpha }^{\sigma } \right] \nonumber \\
& -\frac{\partial }{\partial {{r}_{\alpha }}}\left[ {{\kappa }^{\sigma }}\frac{\partial }{\partial {{r}_{\alpha }}}\left( \frac{D+{{I}^{\sigma }}}{2}\frac{{{T}^{\sigma }}}{{{m}^{\sigma }}} \right)-u_{\beta }^{\sigma }P_{\alpha \beta }^{\sigma }+X_{\alpha }^{\sigma }\right] \nonumber \\
& ={{\rho }^{\sigma }}u_{\alpha }^{\sigma }{{a}_{\alpha }}-\frac{{{\rho }^{\sigma }}}{{{\tau }^{\sigma }}}\left( \frac{D+{{I}^{\sigma }}}{2}\frac{{{T}^{\sigma }}-T}{{{m}^{\sigma }}}+\frac{{{u}^{\sigma 2}}-{{u}^{2}}}{2} \right)
\tt{,}
\end{eqnarray}
with
\begin{equation}
P_{\alpha \beta }^{\sigma }=-{{\mu }^{\sigma }}\left( \frac{\partial u_{\alpha }^{\sigma }}{\partial {{r}_{\beta }}}+\frac{\partial u_{\beta }^{\sigma }}{\partial {{r}_{\alpha }}}-\frac{2{{\delta }_{\alpha \beta }}}{D+I^{\sigma}}\frac{\partial u_{\chi }^{\sigma }}{\partial {{r}_{\chi }}} \right)
\tt{,}
\end{equation}%
\begin{equation}
U_{\alpha \beta }^{\sigma }=-{{\rho }^{\sigma }}\left( {{\delta }_{\alpha \beta }}\frac{{{u}^{\sigma 2}}+{{u}^{2}}-2u_{\chi }^{\sigma }{{u}_{\chi }}}{D+I^{\sigma}}+{{u}_{\alpha }}u_{\beta }^{\sigma }+u_{\alpha }^{\sigma }{{u}_{\beta }}-u_{\alpha }^{\sigma }u_{\beta }^{\sigma }-{{u}_{\alpha }}{{u}_{\beta }} \right)
\tt{,}
\end{equation}%
\begin{equation}
X_{\alpha }^{\sigma }=\frac{{{\rho }^{\sigma }}u_{\alpha }^{\sigma }}{D+I^{\sigma}}{{\left( u_{\beta }^{\sigma }-{{u}_{\beta }} \right)}^{2}}-{{\rho }^{\sigma }}\frac{u_{\alpha }^{\sigma }-{{u}_{\alpha }}}{2}\left[ \frac{D+I^{\sigma}+2}{{{m}^{\sigma }}}\left( {{T}^{\sigma }}-T \right)+{{u}^{\sigma 2}}-{{u}^{2}} \right]
\tt{,}
\end{equation}
where $p^{\sigma }=n^{\sigma}T^{\sigma }$, $e^{\sigma }=(D+I^{\sigma})T^{\sigma }/(2m^{\sigma })$, $\mu ^{\sigma }=p^{\sigma }\tau ^{\sigma }$, $\kappa^{\sigma}=\gamma^{\sigma} \mu ^{\sigma }$, and $\gamma^{\sigma} =(D+I^{\sigma}+2)/(D+I^{\sigma})$ denote the pressure, internal energy per unit mass, dynamic viscosity coefficient, heat conductivity and specific heat ratio of species $\sigma$, respectively.

Applying the operator $\sum_{\sigma }$ to both sides of Eqs. (\ref{NS_sigma_1}) $-$ (\ref{NS_sigma_3}) leads to the NS equations describing the whole system,
\begin{equation}
\label{NS_1}
\frac{\partial \rho }{\partial t}+\frac{\partial }{\partial {{r}_{\alpha }}}\left( \rho {{u}_{\alpha }} \right)=0
\tt{,}
\end{equation}%
\begin{equation}
\label{NS_2}
\frac{\partial }{\partial t}\left( \rho {{u}_{\alpha }} \right)+\frac{\partial }{\partial {{r}_{\beta }}}\sum\nolimits_{\sigma }{\left( {{\delta }_{\alpha \beta }}{{p}^{\sigma }}+{{\rho }^{\sigma }}u_{\alpha }^{\sigma }u_{\beta }^{\sigma } \right)}+\frac{\partial }{\partial {{r}_{\beta }}}\sum\nolimits_{\sigma }{ P_{\alpha \beta }^{\sigma }+U_{\alpha \beta }^{\sigma }}=\rho {{a}_{\alpha }}
\tt{,}
\end{equation}%
\begin{eqnarray}
\label{NS_3}
& \frac{\partial }{\partial t}\left[ \rho \left( e+\frac{1}{2}{{u}^{2}} \right) \right]+\frac{\partial }{\partial {{r}_{\alpha }}}\sum\nolimits_{\sigma }{\left[ {{\rho }^{\sigma }}u_{\alpha }^{\sigma }\left( {{e}^{\sigma }}+\frac{1}{2}{{u}^{\sigma 2}} \right)+{{p}^{\sigma }}u_{\alpha }^{\sigma } \right]} \nonumber \\
& -\frac{\partial }{\partial {{r}_{\alpha }}}\sum\nolimits_{\sigma }{\left[ {{\kappa }^{\sigma }}\frac{\partial }{\partial {{r}_{\alpha }}}\left( \frac{D+I^{\sigma}}{2}\frac{{{T}^{\sigma }}}{{{m}^{\sigma }}} \right)-u_{\beta }^{\sigma }P_{\alpha \beta }^{\sigma }+X_{\alpha }^{\sigma } \right]}=\rho {{u}_{\alpha }}{{a}_{\alpha }}
\tt{,}
\end{eqnarray}%
where $e=\sum_{\sigma }\rho ^{\sigma }(e^{\sigma }+u^{\sigma 2}/2)/\rho -u^{2}/2$ is the internal energy of the physical system per unit mass.

\nocite{*}

\bibliography{References}

\end{document}